\documentclass[ALICE,manyauthors]{cernphprep}
\usepackage[comma,square,numbers,sort&compress]{natbib}
\usepackage{hyperref}
\usepackage{lineno}
\usepackage{color}
\usepackage{caption}
\usepackage{multirow}
\usepackage{floatrow}
\newcommand{\fnorm}{\ensuremath{f_{\mathrm{norm}}}\xspace}
\newcommand{\ptt}{\ensuremath{p_{\mathrm{T}}}\xspace}
\newcommand{\fvtx}{\ensuremath{f_{\mathrm{vtx}}}\xspace}

\newcommand{\dd}{\ensuremath{\mathrm{d}}\xspace}
\newcommand{\KP}{\ensuremath{\mathrm{K}^{\pm}\mathrm{\pi}^{\mp}}\xspace}

\newcommand {\pT}{\ensuremath{p_{\mathrm{T}}}}
\newcommand {\ks}{\mathrm{K}^{*0}}
\newcommand {\ph}{\phi}
\newcommand {\kpi}{$\mathrm{K }\pi ~$}

\newcommand {\modrap} {$\left | y \right | ~$}

\newcommand {\dndy}{d$N$/d$y$}
\newcommand {\dndydpT}{d$^{\mathrm 2}N$/d$y$d\pT}

\newcommand {\pp}{pp}

\newcommand {\tev} {TeV}

\newcommand {\gmom} {GeV/$c$}

\newcommand{\Jpsi} {\mbox{J\kern-0.05em /\kern-0.05em$\psi$}}

\begin{document}%

%%%%%%%%%%%%%%%  Title page %%%%%%%%%%%%%%%%%%%%%%%%
\begin{titlepage}
% CERN-EP-2019-248
\PHyear{2019}
\PHnumber{248}      % required, will be obtained from PH
\PHdate{28 October}  % required, will be obtained from PH
%

%%% Put your own title + short title here:
\title{$\mathrm{K}^{*}(\mathrm{\bf 892})^{0}$ and $\mathrm{\phi(\bf 1020)}$ production at midrapidity in pp~collisions~at~$\sqrt{s}$~=~8~TeV}
\ShortTitle{$\mathrm{K}^{*}(\mathrm{892})^{0}$ and $\mathrm{\phi(1020)}$ production at midrapidity in pp collisions at $\sqrt{s}$ = 8 TeV}   % appears on right page headers

%%% Do not change the next lines
\Collaboration{ALICE Collaboration\thanks{See Appendix~\ref{app:collab} for the list of collaboration members}}
\ShortAuthor{ALICE Collaboration} % appears on left page headers, do not change

 \begin{abstract}
 The production of $\mathrm{K}^{*}(\mathrm{892})^{0}$~and $\mathrm{\phi(1020)}$ in \pp~collisions at 
 $\sqrt{s}$ = 8~\tev~was measured~using~Run~1 data collected by the ALICE collaboration at the LHC. 
 The \pT-differential yields {\dndydpT} in the range 0 $< $~\pT~$ < $~20~\gmom~for $\ks$~and 0.4~$<$~\pT~$<$~16 \gmom~for $\ph$~have been measured at midrapidity, \modrap $<$ 0.5. Moreover, improved measurements of the $\ks(892)$ and $\ph(1020)$ at $\sqrt{s}$~=~7~TeV are presented. The collision energy dependence of \pT~distributions, \pT-integrated yields and particle ratios in inelastic pp collisions are examined. The results are also compared with different collision systems.
 The values of the particle ratios are found to be similar to those measured at other LHC energies. In pp 
 collisions a hardening of the particle spectra is observed with increasing energy, but at the same 
 time it is also observed that the relative particle abundances are independent of the collision energy.
 The \pT-differential yields of $\ks$ and $\ph$ in pp collisions at $\sqrt{s}$ = 8 TeV are compared with the expectations of different Monte Carlo event generators.
 \end{abstract}
 \end{titlepage}
 
 \setcounter{page}{2}
 
 \section{Introduction}
 The study of resonances plays an important role in understanding particle production mechanisms. Particle production at LHC energies has both soft and hard-scattering origins. The hard scatterings are perturbative processes and are responsible for production of high-$\pT$ particles, whereas the bulk of the particles are produced due to soft interactions, which are non-perturbative in nature. High-$\pT$ particles originate from fragmentation of jets and their yield can be calculated by folding the perturbative Quantum Chromodynamics (pQCD) calculations for elementary parton-parton scatterings with universal fragmentation functions determined from experimental data~\cite{Collins:1989gx,deFlorian:2014xna,deFlorian:2017lwf}. The production yield of low-$\pT$ particles can not be estimated from the first principles of QCD, hence predictions require phenomenological models in the non-perturbative regime. In this paper, we discuss $\ks(892)$ and $\ph(1020)$ production in pp collisions at $\sqrt{s}$ = 8~\tev. The $\ph(1020)$ meson is a vector meson consisting of strange quarks ($\mathrm{s\overline{s}}$). The  production of $\mathrm{s\overline{s}}$ pairs was found to be significantly suppressed, compared to $\mathrm{u\overline{u}}$ and $\mathrm{d\overline{d}}$ pairs in pp collisions due to the larger mass of the strange quark~\cite{Malhotra:1982yr,Wroblewski:1985sz}. The $\ks(892)$ is a vector meson with a similar mass to the $\ph(1020)$, but differs in strangeness content by one unit, which may help in understanding the strangeness production dynamics. Measurements of particle production in inelastic pp collisions provide input to tune the QCD inspired Monte Carlo (MC) event generators such as EPOS~\cite{Pierog:2013ria}, PYTHIA~\cite{Skands:2014pea} and PHOJET~\cite{Engel:1995yda,Engel:1994vs}. Furthermore, the measurements in inelastic pp collisions at $\sqrt{s}$ = 8~\tev~reported in this paper serve as reference data to study nuclear effects in proton$-$lead (p--Pb) and lead$-$lead (Pb--Pb) collisions.
 
In this article, the \ptt-differential and \ptt-integrated yields and  the mean transverse momenta of $\ks(892)$ and $\ph(1020)$ at midrapidity in pp collisions at $\sqrt{s}$~=~8~TeV are presented. The energy dependence of the \pT~distributions and particle ratios to the yields of charged pions and kaons in pp collisions is examined and discussed. The yields of pions and kaons measured previously by ALICE~\cite{ALICE_piKp_900GeV,Adam:2015qaa,Abelev:2014laa} at $\sqrt{s}$~=~0.9, 2.76 and~7~$\rm{\tev}$  are used to obtain the yields in pp~collisions at  $\sqrt{s}$~=~8~\tev. Moreover, updated measurements of the $\ks(892)$ and $\ph(1020)$ at $\sqrt{s}$~=~7~TeV are presented; our first measurements for that collision system were published in Ref.~\cite{Abelev:2012hy}. These results include an extension of the $\ks(892)$ measurement to high \pT~and an improved re-analysis of the $\ph(1020)$. This measurement has updated track-selection cuts, which are identical to those described for the measurements at $\sqrt{s}$~=~8~TeV, has an improved estimate of the systematic uncertainties, and extends to greater values of $\pT$. Throughout this paper, the results for K$^{*}(892)^{0}$ and  $\mathrm{\overline{K}}^{*}(892)^{0}$ are averaged and denoted by the symbol $\ks$, while $\phi(1020)$ is  denoted by $\phi$ unless specified otherwise. 
 
This article is organized as follows. The experimental setup is briefly explained in Sec.~\ref{sec2} and the analysis procedure is given in Sec.~\ref{sec3}. The results and discussions are presented in Sec.~\ref{sec4} followed by the conclusions in Sec.~\ref{sec5}.

 \section{Experimental setup}
 \label{sec2}

The ALICE detector can be used to reconstruct and identify particles over a wide momentum range, thanks to the low material budget, the moderate magnetic field (0.5 T) and the presence of detectors with excellent particle identification (PID) techniques. A comprehensive description of the detector and its performance during Run 1 of the LHC is reported in Refs.~\cite{Aamodt:2008zz, Abelev:2014ffa}.

The detectors used for this analysis are described in the following. The V0 detectors are two plastic scintillator arrays used for triggering and event characterization. They are placed along the beam direction at $3.3$ m (V0A) and $-0.9$ m (V0C) on either side of the interaction point with a pseudorapidity coverage of $2.8<\eta<5.1$ and $-3.7<\eta<-1.7$, respectively. The Inner Tracking System (ITS), which is located between 3.9~cm and 43~cm radial distance from the beam axis, is made up of six layers of cylindrical silicon detectors (2 layers of silicon pixels, 2 layers of silicon drift and 2 layers of double-side silicon strips). As it provides high-resolution space points close to the interaction point, the momentum and angular resolution of the tracks reconstructed in the Time Projection Chamber (TPC) is improved.  The TPC is the main tracking device covering full azimuthal acceptance and the pseudorapidity range $-0.9<\eta<0.9$. It is a 92~$\mathrm{m}^3$ cylindrical drift chamber filled with an active gas. It is divided into two parts by a central cathode and the end plates consist of multi-wire proportional chambers. The TPC is also used for particle identification via the measurement of the specific ionization energy loss ($\mathrm{d}E/\mathrm{d}x$) in the gas. The Time of Flight (TOF) detector surrounds the TPC and consists of large multigap resistive plate chambers. It has pseudorapidity coverage  $-0.9<\eta<0.9$, full azimuthal acceptance and an intrinsic time resolution of $<$ 50 ps. The TOF is used for particle identification at intermediate momenta. The particle identification techniques based on the TPC and TOF signals are presented in detail in the next section.

%%%%%%%%%%%%%%%%// Fig. 1 //%%%%%%%%%%%%%%%%%
\section{Data analysis}
\label{sec3}
The measurements of $\ks$ and $\ph$ meson production in pp collisions at $\sqrt{s}$ = 8~\tev~(7 \tev)~were performed during Run 1 data taking with the ALICE detector in 2012 (2010) using a minimum bias trigger as discussed in Sec.~\ref{sec3.1}. A total of around 45M events were analysed for both $\sqrt{s}$ = 7 and 8 TeV and the corresponding integrated luminosities are 0.72 $\rm{nb^{-1}}$ and 0.81 $\rm{nb^{-1}}$, respectively.  The $\ks$ and $\ph$ resonances are reconstructed via their hadronic decay channels with large branching ratios (BR): $\ks$ $\rightarrow$~$\pi^{\pm}$$\mathrm{K}^{\mp}$ with BR = 66.6$\%$ and $\ph$ $ \rightarrow$ $\mathrm{K}^{+}\mathrm{K}^{-}$ with BR = 49.2$\%$~\cite{Tanabashi:2018oca}. Some older measurements of $\ph$ used a value of 48.9\% for the $\ph$ $ \rightarrow$ $\mathrm{K}^{+}\mathrm{K}^{-}$ branching ratio~\cite{Patrignani:2016xqp}; when comparing different $\ph$ measurements, the older results are scaled to account for the new branching ratio.

\subsection{Event and track selection}
\label{sec3.1}
For pp collisions at $\sqrt{s}$ = 8~\tev, the events were selected with a minimum bias trigger based on a coincidence signal in V0A and V0C. For pp collisions at $\sqrt{s}$ = 7~\tev, the trigger condition is same as in~\cite{Abelev:2012hy}. The ITS and TPC are used for tracking and reconstruction of charged particles and of the primary vertex. Events having the primary vertex coordinate along the beam axis within 10 cm from the nominal interaction point are selected. Pile-up events are rejected if  more than one vertex is found with the Silicon Pixel Detector (SPD). A primary track traversing the TPC induces signals on a maximum of 159 tangential pad-rows, each corresponding to one cluster used in track reconstruction. For this analysis high quality charged tracks are used to select pion and kaon candidates coming from the decays of $\ks$ and $\phi$. Tracks are required to have at least 70 TPC clusters and a $\chi^{2}$ per track point ($\chi^{2}/N_{\mathrm{clusters}}$) of the track fit in the TPC less than 4. 
Moreover, tracks must be associated with at least one cluster in the SPD. To ensure a uniform acceptance by avoiding the edges of the TPC, tracks are selected within $|\eta|$~$\textless$~0.8. In order to reduce contamination
from secondary particles coming from weak decays, 
cuts on the distance of closest approach to the primary vertex in the transverse plane 
(DCA$_{xy}$) and longitudinal direction (DCA$_z$) are applied. The value of DCA$_{xy}$ 
is required to be less than 7 times its resolution: DCA$_{xy}(p_{\mathrm{T}}) 
~\textless ~ (0.0105 + 0.035p_{\mathrm{T}}^{-1.1}$)~cm ($p_{\mathrm{T}}$ in GeV/$c$) 
and DCA$_{z}$ is required to be less than 2 cm.  To improve the global resolution, the $p_{\mathrm{T}}$ of each track 
is choosen to be greater than 0.15 GeV/$c$. 

In the TPC, particles are identified by measuring the d$E$/d$x$ in the TPC gas, whereas in the TOF it 
is done by measuring the time of flight. The particles in the TPC are selected using a cut on 
the difference of the mean value of the d$E$/d$x$ to the
expected d$E$/d$x$ value for a given species divided by the resolution $\sigma_{\rm{TPC}}$. This cut is expressed in units of the estimated $\sigma_{\rm{TPC}}$. As described below, this is optimized for each analysis and depends on the signal-to-background 
ratio and on the transverse momentum. Particles are identified in the TOF by comparing
the measured time of flight to the expected one for a given particle species. The cut is 
expressed in units of the estimated resolution $\sigma_{\rm{TOF}}$. The TOF allows pions and kaons to be unambiguously identified 
up to momentum $ p \approx 1.5 ~\mathrm{GeV}/c$ and also removes contamination from electrons. The two mesons can be distinguished from (anti)protons up 
to $p$ $\approx$ 2.5 GeV/$c$.

For $\ks$ and $\ph$ reconstruction three TPC PID selection criteria are used, depending on the momentum of the
daughter particle. For pp collisions at $\sqrt{s}$ = 8 TeV, both pions and kaons are selected using a cut of 
$\mid \rm{N}\sigma_{\mathrm{TPC}}\mid < 2.0$ for $p(K^\pm,\pi^\pm)~>~0.4~\mathrm{GeV}/c$. Here, $p(K^\pm,\pi^\pm)$ denotes the momenta of pions and kaons. Similarly, for $p(K^\pm,\pi^\pm)~<~0.3$~GeV/$c$, a cut of $|\rm{N}\sigma_{\mathrm{TPC}}|~<~6.0$
is applied, while a cut of $|\rm{N}\sigma_{\mathrm{TPC}}|~<4.0$~for~$ 0.3~<p(K^\pm,\pi^\pm)~<~0.4~\mathrm{GeV}/c ~$
is applied. For the new analysis of the $\ks$ ($\ph$) at $\sqrt{s}$ = 7 TeV, the specific energy loss for pion and kaon candidates is required to be within 2 (3) $\sigma_{\rm{TPC}}$ of the expected mean, irrespective of the momentum. Also, a TOF~3$\sigma_{\rm{TOF}}$~veto cut is applied for $\ks$ for both $\sqrt{s}$ = 7 and 8 TeV. ``TOF 
veto" means that the TOF  3$\sigma$ cut is applied only for cases where the track matches a hit in 
the TOF.

\subsection{Raw yield extraction} \label{sec:signal}
\begin{figure}[ht]
  \begin{center}
 \includegraphics[width=7cm, height=6cm]{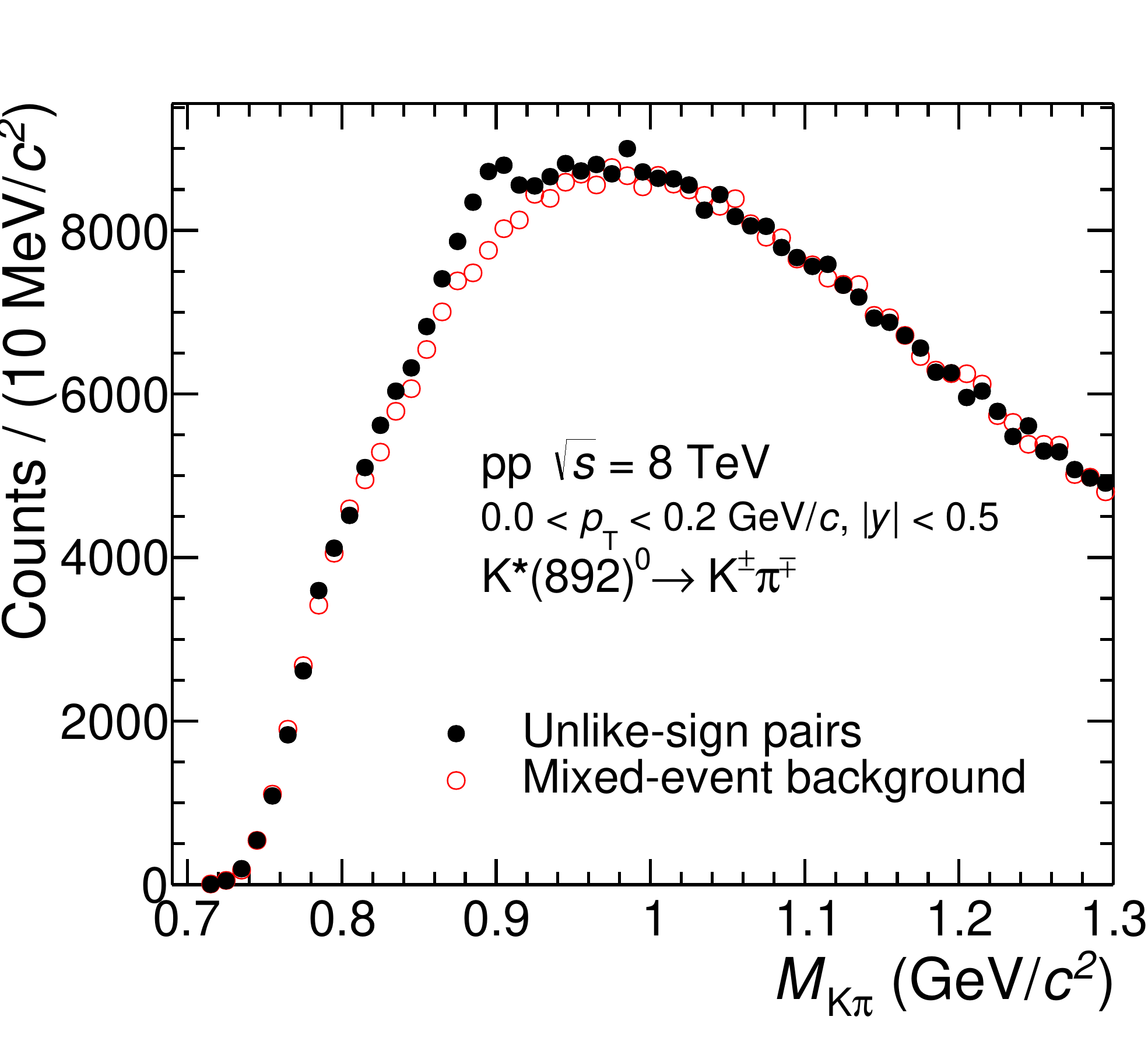}
 \includegraphics[width=7cm, height=6cm]{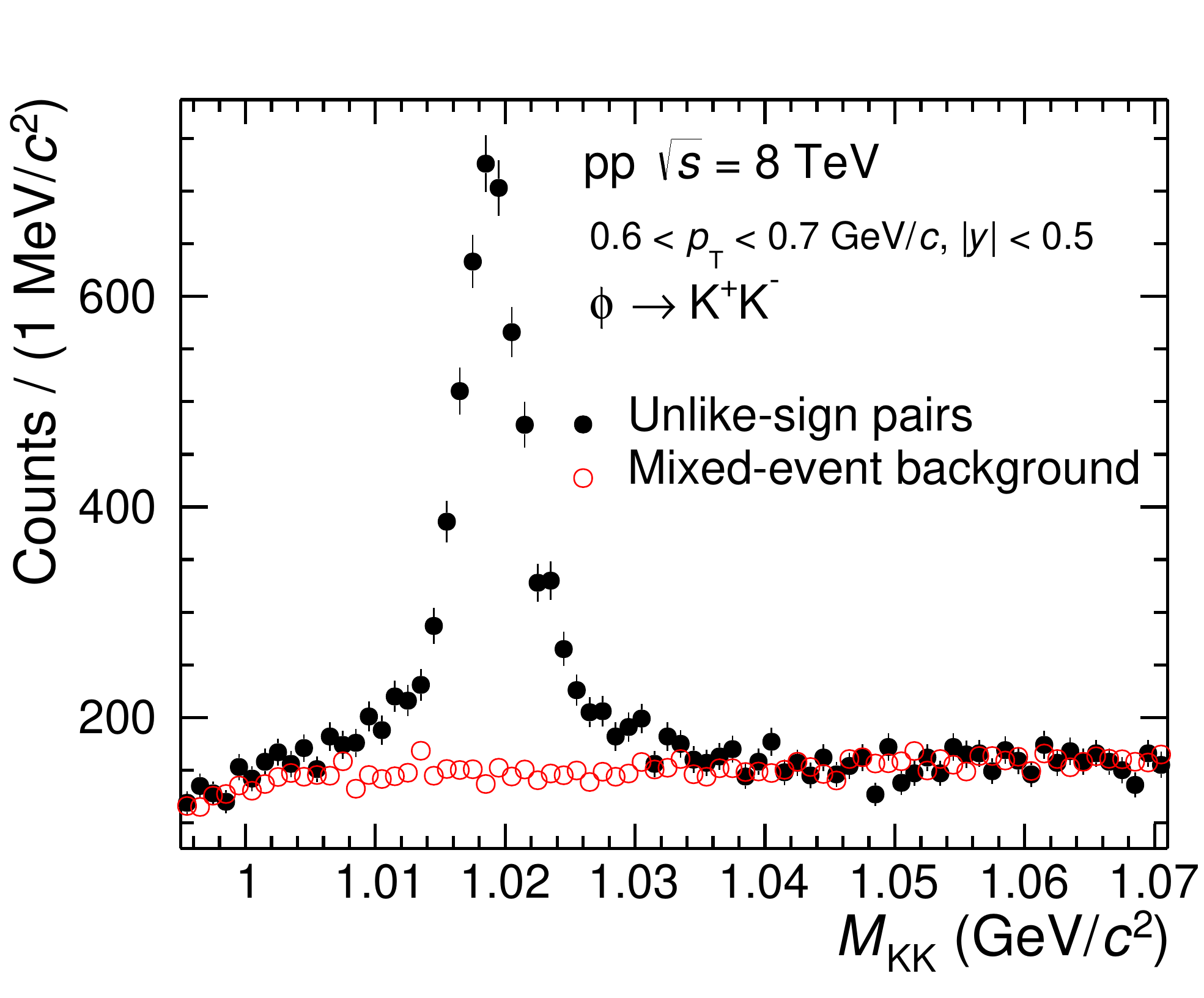}
\includegraphics[width=7cm, height=6cm]{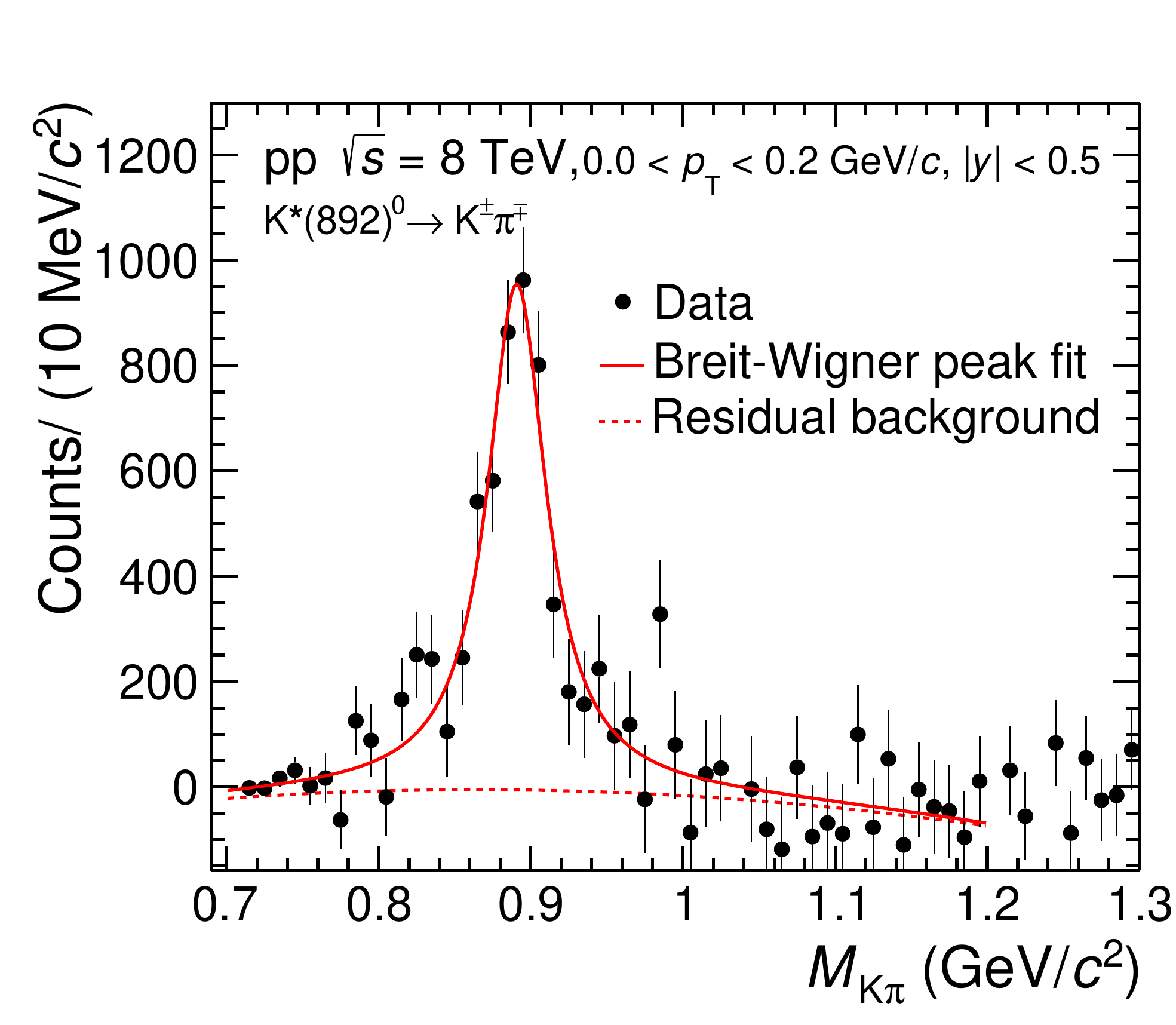}
 \includegraphics[width=7cm, height=6cm]{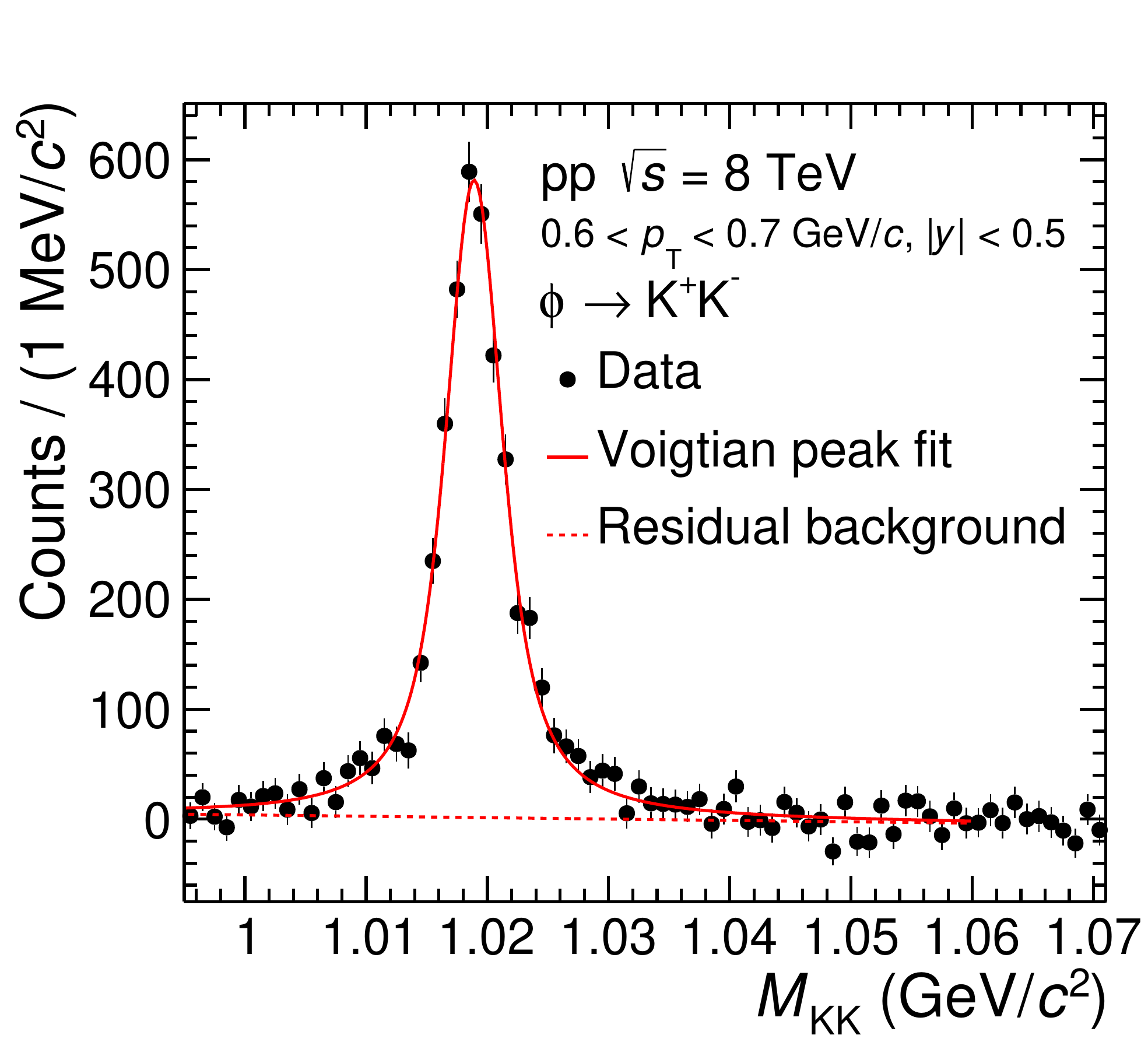}
 \caption{(Color online)  (Upper panels) Invariant mass distributions (closed black point) for the $\ks$ (left) and $\ph$ (right) 
 in pp collisions at 8 TeV in the $\pT$ range  0 $< $~\pT~$<$~0.2 GeV/$c$ and 0.6 $< $~\pT~$<$~0.7 GeV/$c$, respectively. The combinatorial background (open red circles) is estimated using unlike-sign pairs from different events 
 (mixed event). The statistical uncertainties are shown as bars. (Lower panels) \kpi (left) and KK (right) invariant mass distributions in the same \ptt ranges after combinatorial background subtraction together with the fits to the signal and background contribution.}
    \label{pp_8:Fig1} 
  \end{center}
\end{figure}

The $\ks$ ($\phi$) meson is reconstructed through its dominant hadronic decay 
channel $\ks \rightarrow \pi^{\pm} \mathrm{K}^{\mp} ~\mathrm~ (\ph  \rightarrow  ~\mathrm{K}^{+}\mathrm{K}^{-}$)
 by calculating the invariant mass of its daughters at the primary 
vertex. The invariant mass distribution of the decay daughter pairs is constructed by
taking unlike-sign pairs of K and 
$\pi$ (K) candidates for $\ks$ ($\phi$) in the same event. The rapidity of the $\pi$K (KK) 
pairs is required to lie within the range $|y_{\mathrm{pair}}|$ $\textless$ 0.5. As an example, the $\pi$K (KK) invariant mass distribution for $\sqrt{s}$~=~8~TeV is shown in Fig.~\ref{pp_8:Fig1} for $0 < \pT < 0.2$~\gmom~($0.6~<~\pT~<~0.7$~\gmom).

The shape of the uncorrelated background is obtained via the event mixing technique, calculating the invariant mass distribution of unlike-sign $\pi^{\pm}$$\mathrm{K}^{\mp}$ ($\ks$) or $\mathrm{K}^{+}\mathrm{K}^{-}$ ($\ph$) combinations from different events, as shown in the upper panel of Fig.~\ref{pp_8:Fig1}. To reduce statistical uncertainties each event was mixed with 5 other similar events. For $\sqrt{s}$~=~8~TeV, the mixed event background is normalized in the mass range $1.1< $ $M_{\rm{K}\pi }$ $< 1.5~\mathrm{GeV}/c^{2}$   ($1.04< $ $M_{\rm{KK}}$ $<1.06~\mathrm{GeV}/c^{2}$) for $\ks$($\ph$) so that it has the same integral as the unlike-charge 
  distribution in that normalization region. For $\sqrt{s}$~=~7~TeV, the mixed event background is normalized in the mass range $1.1< $ $M_{\rm{K}\pi }$ $< 1.15~\mathrm{GeV}/c^{2}$ and $1.048< $ $M_{\rm{KK}}$ $<1.052~\mathrm{GeV}/c^{2}$ for $\ks$ and $\ph$, respectively. To avoid mismatches due to different acceptances and to assure a similar event structure, only tracks from events with similar vertex positions ($\Delta z$ $<$ 1 cm) and track multiplicities ($\Delta n$ $<$ 5) are mixed. For the $\phi$ meson in pp collisions at $\sqrt{s}$~=~7~TeV, the multiplicity difference for event mixing is restricted to $\Delta n$ $\leq$ 10. This combinatorial background is subtracted from the 
  unlike-charge mass distribution  in each $\pT$ bin.  Due to an imperfect description of the combinatorial 
  background, as well to the presence of a correlated background, a residual
  background still remains. The correlated background can arise from correlated
  \kpi (KK) pairs for $\ks$($\ph$), misidentified particle decays, or jets.

 The $\ks$ raw yield is extracted  from the \kpi  invariant mass distribution in different \ptt bins between
 0 and 20 GeV/$c$. After the combinatorial background subtraction the invariant mass distribution 
 is  fitted with the combination of a Breit-Wigner function for the signal peak and a second-order polynomial for the residual background.
 The fit function for $\ks$ is given by

\begin{equation}
\label{eq:kstar:signal}
\frac{\dd N}{\dd M_{\mathrm{\KP}}}=\frac{A}{2\pi} \times \frac{\Gamma_0} {(M_{\KP}-m_0)^2+
\frac{\Gamma_0^2}{4}}+(BM_{\mathrm{\KP}}^{2} + CM_{\mathrm{\KP}} + D).
\end{equation}

Here $m_{\mathrm{0}}$ is the fitted mass pole of the $\ks$, $\Gamma_{\mathrm{0}}$ 
is the resonance width and $A$ is the
yield of the $\ks$ meson. $B$, $C$ and $D$ are the 
fit parameters in the second-order polynomial.

The $\ph$ raw yield is extracted from the KK invariant mass distribution in different \ptt bins between 0.4 and 16 GeV/$c$ after
the combinatorial background subtraction. For the $\ph$ fit function, the detector mass 
resolution is taken into account due to the smaller width of the $\ph$ meson. This is achieved by using a 
Breit-Wigner function convoluted with a Gaussian function, which is known as Voigtian function.
The KK invariant mass distribution is fitted with the combination of a Voigtian function for the signal peak 
and a second-order polynomial for the residual background. The fit function for $\phi$ is given by
 
 \begin{equation}
  \frac{\mathrm{d}N}{\mathrm{d}M_{\mathrm{KK}}} = \frac{A \Gamma_{0}}
       {(2\pi^{3/2})\sigma} \times \int\limits_{-\infty}^{+\infty} \mathrm{exp} 
       \Bigg(\frac{(M_{\mathrm{KK}} - m')^{2}}{2\sigma^{2}}\Bigg) 
       \frac{1}{(m'-m_{0})^{2} + \frac{\Gamma_{0}^{2}}{4} } \mathrm{d}m' 
       + (BM_{\mathrm{KK}}^{2} + CM_{\mathrm{KK}} + D).
       \label{eq2}
\end{equation}
Here $m_{\mathrm{0}}$ is the fitted mass pole of the $\ph$, $\Gamma_{\mathrm{0}}$ 
is the resonance width fixed to the value in vacuum and $\sigma$ is the $p_{\mathrm{T}}$-dependent mass resolution, which ranges from 1 to 3 MeV/$\it{c^2}$.

To extract the raw yields of $\ks$ ($\phi$), for each $p_{\mathrm{T}}$ bin the invariant mass histogram is integrated over the region 
$0.801<m_{\ks}<0.990~(1.01<m_{\ph}<1.03$), i.e. a range of 2-3 times the nominal width around the nominal mass. The integral of the residual background function in the same range is then subtracted.
The resonance yields beyond the histogram integration regions are found by integrating the tails of the signal fit function; these yields are then added to the peak yield computed by integrating the histogram.

\subsection{Normalization and correction}
The $\ks$ and $\ph$ raw yields ($N_{\mathrm{raw}}$) are normalized to the number of inelastic pp collisions and corrected for the 
branching ratio (BR), vertex selection, detector geometric acceptance (A) and efficiency (\ensuremath{\varepsilon}) and signal loss. The $\ks$ and $\ph$ corrected yields are obtained by 

\begin{equation}
\dfrac{\dd^{2}N}{\dd\ptt\dd y}=\dfrac{N_{ \mathrm{raw} } \times \epsilon_{\mathrm{SL}}} {N_{\mathrm{evt}}\times 
\mathrm{BR}\times\dd\ptt  \times  \dd y \times \ensuremath{\varepsilon_{\mathrm{rec}}}}\times\fnorm\times\fvtx  .
\end{equation}

Here \ensuremath{\varepsilon_{\mathrm{rec}}} = $A \times \ensuremath{\varepsilon}$ is the correction that accounts for the detector acceptance and efficiency.
The $\epsilon_{\mathrm{SL}}$ is the signal loss correction factor and accounts for the loss of $\ks (\ph)$ 
mesons incurred by selecting events that satisfy only the ALICE minimum bias trigger, rather than all inelastic events. This is a particle species and $\pT$-dependent correction factor which is peaked at low $\pT$, indicating that events that fail the trigger selection have softer 
\ptt spectra than the average inelastic event. The signal loss correction factor is about 1\% at low-\pT~and negligible for $\pT > $ 1 GeV/$c$. This correction is the ratio of the $\pT$ spectrum from inelastic 
events to the $\pT$ spectrum from triggered events and it is evaluated using Monte Carlo simulations.

$N_{\mathrm{evt}}$ is the number of triggered events and a trigger efficiency (${\it \fnorm}$) is used to normalize the yield to the number of inelastic pp collisions. 
The value of the inelastic normalization  factor for pp collisions at $\sqrt{s}$~=~8 TeV is 0.77 $\pm$ 0.02, which is the ratio between the V0 visible cross section~\cite{ALICE-PUBLIC-2017-002} and the inelastic cross section~\cite{Loizides:2017ack}. Similarly, we correct the yield with \fvtx, which is the ratio of  the number of events for which a good vertex was found to the total number of triggered events. This is estimated to be 0.972. The new results at 7 TeV are normalized as in~\cite{Abelev:2012hy}.

The  \ensuremath{\varepsilon_{\mathrm{rec}}} correction factor is determined with a Monte Carlo simulation using PYTHIA8 as the event generator and GEANT3~\cite{Brun:1978fy} as the transport code for the simulation of the detector response. The \ensuremath{\varepsilon_{\mathrm{rec}}} is obtained as the fraction of $\ks$ 
and $\ph$ reconstructed after passing the same event selection and track quality cuts as used for the real events to the total number of generated resonances. This  \ensuremath{\varepsilon_{\mathrm{rec}}} value is small at low $\pT$ and increases with increasing $\pT$. This value is independent of $\pT$ above 5-6 GeV/$c$~\cite{Abelev:2012hy}.

\subsection{Systematic uncertainties}

The systematic uncertainties on the \ptt-differential yield, summarised in Table~\ref{tab_systematic}, are due to different sources such as signal extraction, background subtraction, track selection, global tracking uncertainty, knowledge of the material budget and the hadronic interaction cross section.  

 \begin{table}
      \begin{center}
          \begin{tabular}{ccccc}
            \hline \hline
            \multicolumn{1}{c}{}&
            \multicolumn{2}{c}{pp, $\sqrt{s} = 8$~TeV} &
            \multicolumn{2}{c}{pp, $\sqrt{s} = 7$~TeV}\\
            \hline
            Source                   &  $\ks$ ($\%$) &  $\phi$ ($\%$ ) &  $\ks$ ($\%$) &  $\phi$ ($\%$ )\\
            \hline 
            Signal extraction                       &   8.7      	&   1.9         &   8.5      	&   4.0   \\
            Track selection                          &   4.0   		&   2.0         &   5.8   		&   3.2 \\
            Material budget                         &  0 -- 3.4     	&   0 -- 5.4   &  0 -- 3.4   	&   0 -- 5.4  \\
            Hadronic Interaction		   &  0 -- 2.8       &   0 -- 3.1  &  0 -- 2.8    	&   0 -- 3.1 \\
            Global tracking efficiency          &  6.0   		&   6.0         &  8.0   		&   8.0 \\
            Branching ratio                          & neg.		&  1.0          & neg.	         &  1.0\\
            \hline
            Total                                          & 11.3 -- 12.1     &  6.7 -- 9.1 & 9.2 -- 18.3     &  9.1 -- 15.4  \\
            \hline \hline

          \end{tabular}
        
      \end{center}
        \caption{Systematic uncertainties in the measurement of $\ks$ and $\phi$ 
        yields in pp collisions at $\sqrt{s}$ = 7 and 8 TeV. 
        The global tracking uncertainty is $p_{\mathrm{T}}$-independent, while 
        the other single-valued systematic uncertainties are averaged over 
        $p_{\mathrm{T}}$. The values given in ranges are minimum and maximum 
        uncertainties depending on $p_{\mathrm{T}}$.
      }
      \label{tab_systematic}
    \end{table}

The systematic uncertainties associated to the signal extraction are estimated by varying the fitting ranges, the order of residual backgrounds (from 1\textsuperscript{st} order to 3\textsuperscript{rd} order), the width parameter and the mixed event background normalization range. The signal extraction systematic uncertainties also include the background subtraction systematic uncertainties, which are estimated by changing the methods used to estimate the combinatorial background (like-sign and event-mixing). The PID cuts and the track quality selection criteria are varied to obtain the systematic uncertainties due to the track selection. The relative uncertainties due to signal extraction and track selection for $\ks$ ($\ph$) are 8.7\% (1.9\%) and 4\% (2\%), respectively at $\sqrt{s}$~=~8~TeV.

The global tracking uncertainty is calculated using ITS and TPC clusters for charged decay daughters. The relative systematic uncertainty due to the global tracking efficiency is 3$\%$ for charged particles, which results in a 6$\%$ effect for the $\pi \rm{K}$ and $\mathrm{KK}$ pairs used in the reconstruction of the $\ks$ and $\ph$, respectively. The systematic uncertainty due to the residual uncertainty in the description of the material in the Monte Carlo simulation contributes up to 3.4$\%$ for $\ks$ (5.4\% for $\ph$). The systematic uncertainty due to the hadronic interaction cross section in the detector material is estimated to be up to 2.8\% for $\ks$ and up to 3.1\% for $\ph$. The uncertainties are accordingly propagated to the $\ks$ and $\ph$~\cite{Abelev:2013vea, Abelev:2013haa}. The total systematic uncertainties, which are found to be  $\pT$ dependent, range in from 11.3\% to 12.1\% for $\ks$ and from 6.7\% to 9.1\% for $\ph$. The uncertainties at $\sqrt{s}$ = 7 TeV are similarly estimated, totalling to comparable values, as seen in Table~\ref{tab_systematic}.

%\clearpage

\section{Results and discussion}
\label{sec4}

\subsection{Transverse momentum spectra and differential yield ratios}
\begin{figure}[!ht]
  \begin{center}
   \includegraphics[scale=0.35]{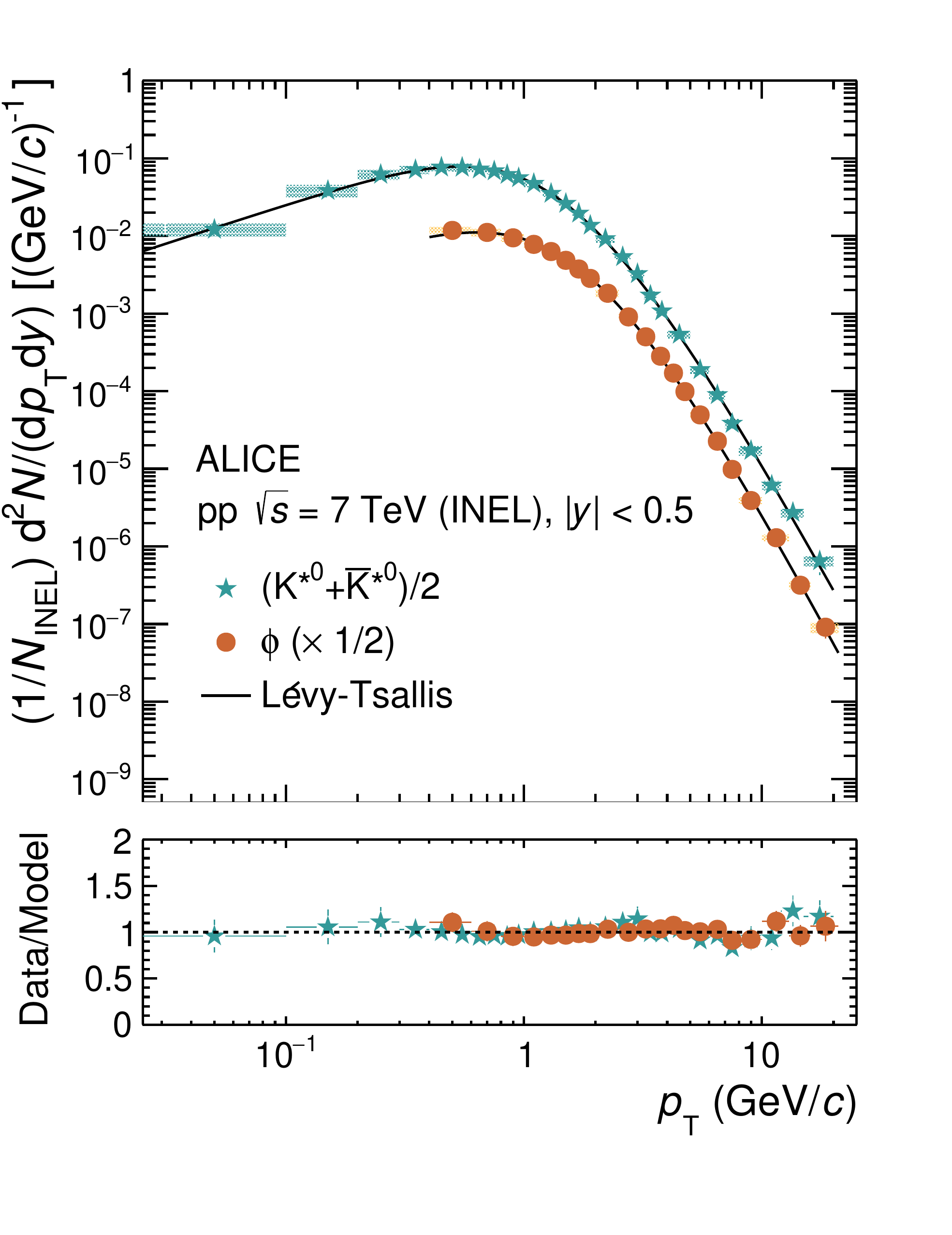}
   \includegraphics[scale=0.35]{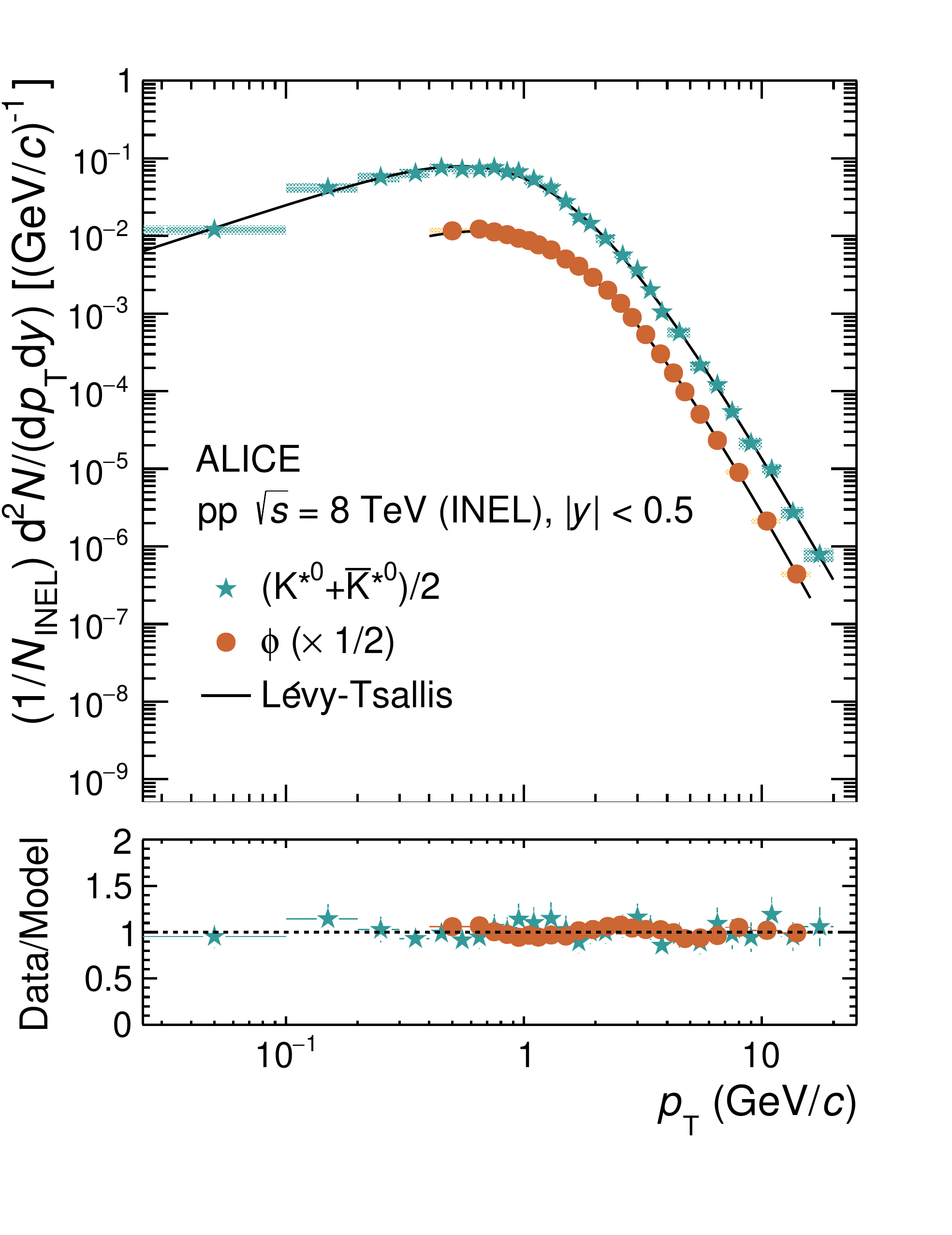}
    \caption{ (Color online) Upper panels shows the $\pT$ spectra of $\ks$ and $\ph $ in inelastic pp collisions at 7 TeV (left) and 8 TeV (right) and fitted with the L\'evy-Tsallis distribution~\cite{Tsallis,STAR_strange_pp_2007}. The normalization uncertainty in the spectra is $^{+7.3}_{-3.5}$\% for 7 TeV and 2.69\% for 8 TeV. The vertical bars show statistical and the boxes show systematic uncertainties. The lower panels show the ratio of data to the L\'evy-Tsallis fit. Here, the bars show the systematic uncertainty.}
    \label{pp_8:Fig2}
  \end{center}
\end{figure}

Here, we report the measurement of $\ks$ and $\ph$ in inelastic pp collisions at $\sqrt{s}$ = 8 TeV in the range up to \pT~$=$~20~\gmom~for $\ks$ and up to \pT~$=$~16~\gmom~for $\ph$. Also, we present the new measurements of  $\ks$ and $\ph$ in inelastic pp collisions at $\sqrt{s}$ = 7 TeV in the range up to \pT~$=$~20~\gmom~for $\ks$ and up to \pT~$=$~21~\gmom~for $\ph$. The re-analyzed $\ks$ and $\ph$ spectra in pp collisions at $\sqrt{s}$ = 7 TeV agree with the previously published values~\cite{Abelev:2012hy} within a few percent at low \pT. At higher \pT~($\gtrsim 3$ \gmom~for $\ks$ and $\gtrsim 2$ \gmom~for $\ph$), the old and re-analyzed results can differ by up to 20\%, although their systematic uncertainties still overlap.
For both energies, the first bin of $\ks$ starts at \pT~$=$~0~\gmom~and for $\ph$, it starts at \pT~$=$~0.4~\gmom. In Fig.~\ref{pp_8:Fig2}, we show the transverse momentum spectra of $\ks$ and $\ph$ at midrapidity \modrap $<$ 0.5 and fitted with the L\'evy-Tsallis distribution~\cite{Tsallis,STAR_strange_pp_2007}. The ratio of the measured data to the L\'evy-Tsallis fit shows good agreement of data with model within systematic uncertainties. The fit parameters are shown in Table~\ref{tab_tsallisfit}. 

\begin{figure}[ht]
  \begin{center}
   
   \includegraphics[width=7cm, height=7cm]{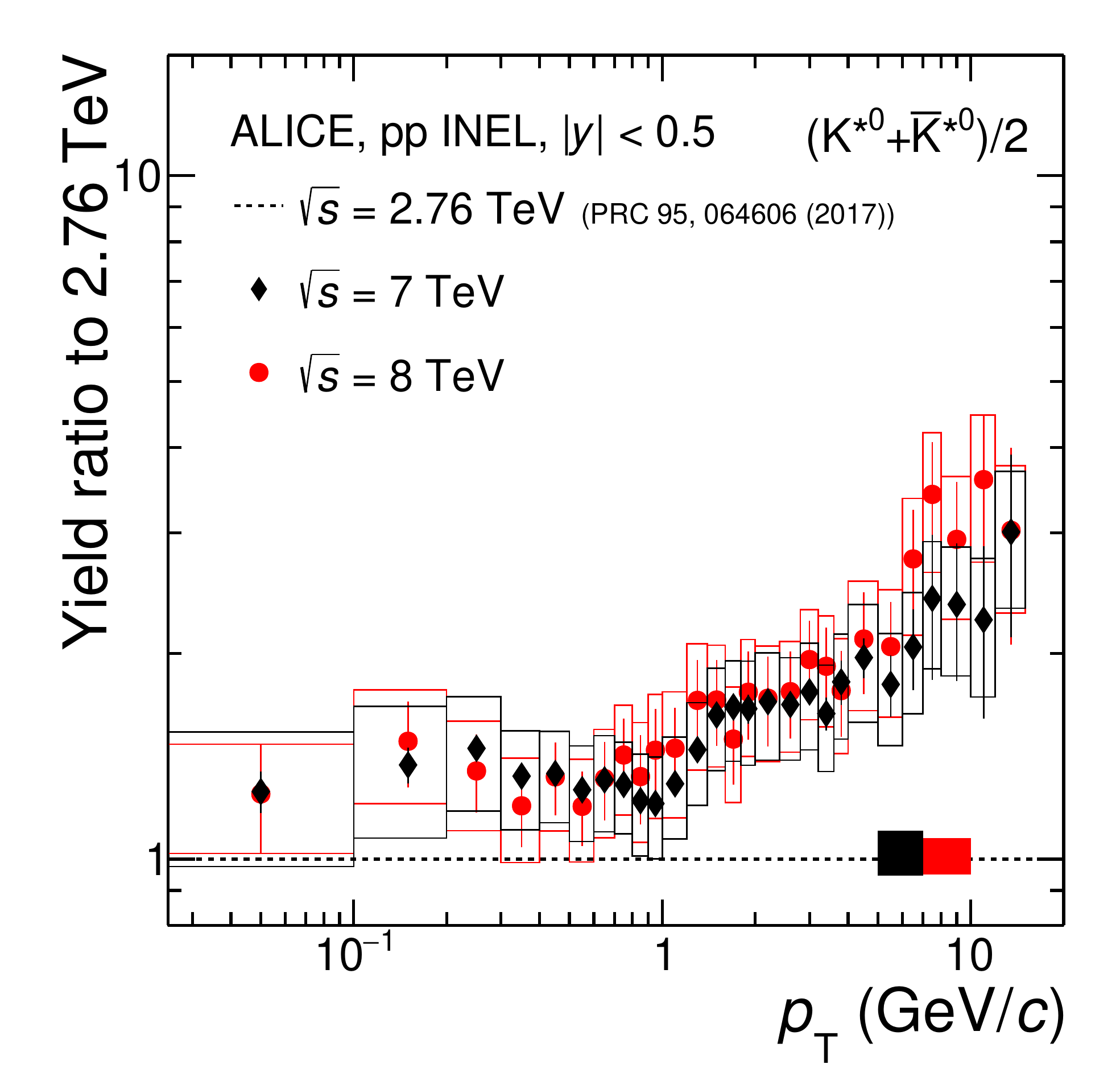}
 \includegraphics[width=7.0cm, height=7.cm]{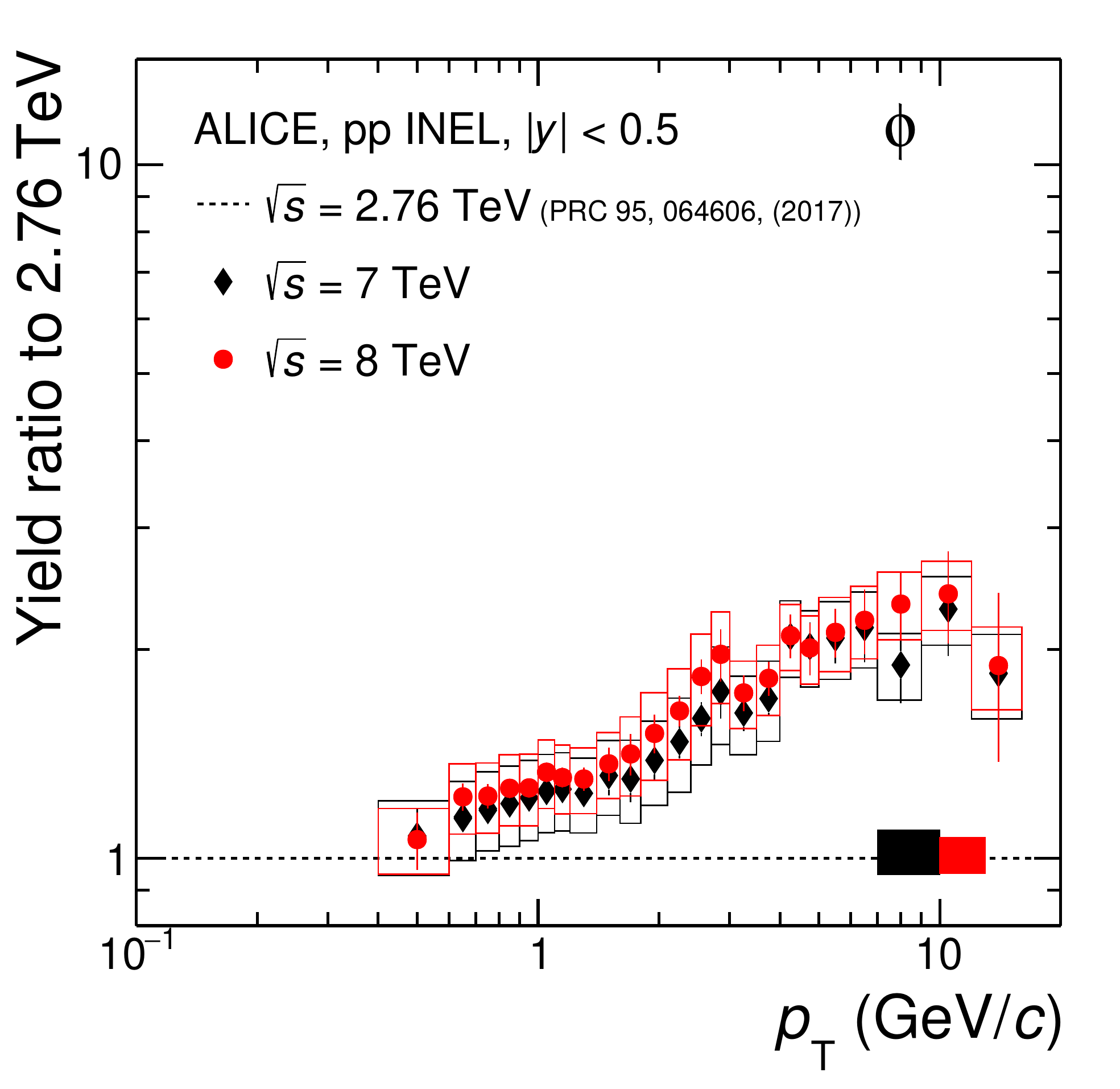}
    \caption{(Color online) Ratios of transverse-momentum spectra of $\ks$ and $\ph $ in inelastic events
     at $\sqrt{s}$~=~7 and 8 TeV to the transverse-momentum spectra in pp collisions at $\sqrt{s}$~=~2.76 TeV. The statistical and systematic uncertainties are shown as vertical error bars and boxes, respectively. The normalization uncertainties are indicated by boxes around unity.}
    \label{pp_8:Fig3}
  \end{center}
\end{figure}

   The energy evolution of the transverse momentum spectra for $\ks$ and $\ph$ is studied by calculating the ratio of \pT-differential 
yields for inelastic events at $\sqrt{s}$~=~7 and 8 TeV to those at $\sqrt{s}$ = 2.76 TeV~\cite{Adam:2017zbf}. This is shown in Fig.~\ref{pp_8:Fig3}. The differential yield ratio to 2.76 TeV is consistent for 7 and 8 TeV within systematic uncertainties. The systematic uncertainties at both collision energies are largely uncorrelated. Therefore, the sum of these in quadrature is taken as systematic uncertainty on the ratios. For both $\ks$ and $\ph$, the differential yield ratio is independent of $\pT$ within systematic uncertainties up to about 1 GeV/$c$ for the different collision energies. This suggests that the particle production mechanism in soft scattering regions is independent of collision energy over the measured energy range. An increase in slope of the differential yield ratios is observed for $\pT > $ 1-2 GeV/$c$. 

  \begin{table}
      \begin{center}
        \scalebox{1.0}{
          \begin{tabular}{ccccc}
          \hline
            \multicolumn{1}{c}{}&
            \multicolumn{2}{c}{pp, $\sqrt{s} = 8$~TeV} &
            \multicolumn{2}{c}{pp, $\sqrt{s} = 7$~TeV}\\
            \hline
            Particles          & T (MeV)        &  n   & T (MeV)        &  n		\\
            \hline 
            $\ks$               & 260 $\pm$ 5 &   6.65 $\pm$ 0.03 & 261 $\pm$ 6 &   6.92 $\pm$ 0.15 	\\
            \hline
            $\ph$           	  & 306 $\pm$ 6 &   7.28 $\pm$ 0.03 & 299 $\pm$ 5 &   7.17 $\pm$ 0.04  \\
            \hline
          \end{tabular}
        }
      \end{center}
	\caption{Parameters extracted from the L\'evy-Tsallis fit to the $\ks$ and $\ph$ transverse momentum spectra in inelastic pp collisions at $\sqrt{s}$ = 7 and 8 TeV.}
      \label{tab_tsallisfit}
    \end{table}

\subsection{\pT-integrated yields}

  \begin{table}
      \begin{center}
        \scalebox{0.9}{
          \begin{tabular}{ccccc}
            \hline
            \multicolumn{4}{c}{pp, $\sqrt{s} = 8$~TeV}\\
            \hline
            Particles         & measured $\pT$ (GeV/$c$)        &  d$N$/d$y$ 	 								&  $\langle \pT \rangle$ (GeV/$c$)\\
            \hline 
            $\ks$               & 0.0 -- 20.0    			 &   0.101 $\pm$ 0.001 (stat.) $\pm$ 0.014 (sys.)       		&   1.037 $\pm$ 0.006 (stat.) $\pm$ 0.029 (sys.) \\
            \hline
            $\ph$           	& 0.4 -- 16.0				&   0.0335 $\pm$ 0.0003 (stat.) $\pm$ 0.0030 (sys.)                &   1.146 $\pm$ 0.005 (stat.) $\pm$ 0.040 (sys.)   \\
            \hline
             \hline
            \multicolumn{4}{c}{pp, $\sqrt{s} = 7$~TeV}\\
            \hline
            Particles        & measured $\pT$ (GeV/$c$)        &  d$N$/d$y$ 	 								                 &  $\langle \pT \rangle$ (GeV/$c$)\\
            \hline 
            $\ks$            & 0.0 -- 20.0    			 &   0.0970 $\pm$ 0.0004 (stat.) $\pm$ 0.0103 (sys.)       		&   1.015 $\pm$ 0.003 (stat.) $\pm$ 0.030 (sys.) \\
            \hline 
            $\ph$           & 0.4 -- 21.0				&   0.0318 $\pm$ 0.0003 (stat.) $\pm$ 0.0032 (sys.)                    &   1.132 $\pm$ 0.005 (stat.) $\pm$ 0.023 (sys.)   \\
            \hline

          \end{tabular}
        }
      \end{center}
	\caption{$\ks$ and $\ph$ integrated yields and $\langle \pT \rangle$ in inelastic pp collisions at $\sqrt{s}$ = 7 and 8 TeV. The systematic uncertainties include the contributions from the uncertainties listed in Table~\ref{tab_systematic} and the choice of the spectrum fit function for extrapolation is also included for the $\ph$. Here, ``stat." and ``sys." refer to statistical and systematic uncertainties, respectively. In addition, the \dndy~has uncertainties due to normalization, which is $^{+7.3}_{-3.5}$\% for 7 TeV and 2.69\% for 8 TeV.}
      \label{tab_yield}
    \end{table}

    Table~\ref{tab_yield} shows the $\ks$ and $\ph$ integrated yield (d$N$/d$y$) and mean transverse momenta ($\langle \pT \rangle$) in inelastic pp collisions at $\sqrt{s}$ = 8 TeV.  As the $\ph$ spectrum starts from 0.4 GeV/$c$, for the calculation of d$N$/d$y$ and $\langle \pT \rangle$,  the spectrum is extrapolated down to \pT~= 0 GeV/$c$ using a L\'evy\,--\,Tsallis fit~\cite{Tsallis,STAR_strange_pp_2007}. The extrapolated part amounts to about 15\% of the yield. Alternative fit functions (Boltzmann distribution, Bose-Einstein distribution, $m_{\rm{T}}$ exponential and \pT~exponential) have been tried for the extrapolation, giving a contribution of 1.5\% to the total systematic uncertainty on d$N$/d$y$. In the case of $\ks$, no extrapolation is needed as the distribution is measured for \pT~$>$ 0 GeV/$c$. Table~\ref{tab_yield} also shows the d$N$/d$y$ and $\langle \pT \rangle$ of $\ks$ and $\ph$ at $\sqrt{s}$ = 7 TeV.

\subsection{Particle ratios}

%\clearpage

%K*/pi ratio
For the calculation of the particle yield ratios, the values of d$N$/d$y$ for $\pi^{+}+\pi^{-}$ and K$^{+}$+K$^{-}$ in pp collisions at $\sqrt{s}$~=~8 TeV are estimated via extrapolation using the data points available at different LHC collision energies~\cite{ALICE_piKp_900GeV,Adam:2015qaa,Abelev:2014laa} namely 0.9, 2.76 and 7 TeV. The data points are fitted with the following polynomial function, $A(\sqrt{s})^{n} + B$.
Here $A$, $n$ and $B$ are the fit parameters. For the calculation of the uncertainties on the extrapolated value, the central values of the data 
points are shifted within their uncertainties and fitted with the same function. The $\pi^{+}+\pi^{-}$and K$^{+}$+K$^{-}$ energy extrapolated yields in inelastic pp collisions at $\sqrt{s}$ = 8 TeV are 4.80 $\pm$ 0.21 and 0.614 $\pm$ 0.032. From here onwards, $\pi^{+}+\pi^{-}$ is denoted as $\pi$ and  K$^{+}$+K$^{-}$ is denoted as K.
 
\begin{figure}[ht]
  \begin{center}
  \resizebox{0.95\textwidth}{!}{
 \includegraphics{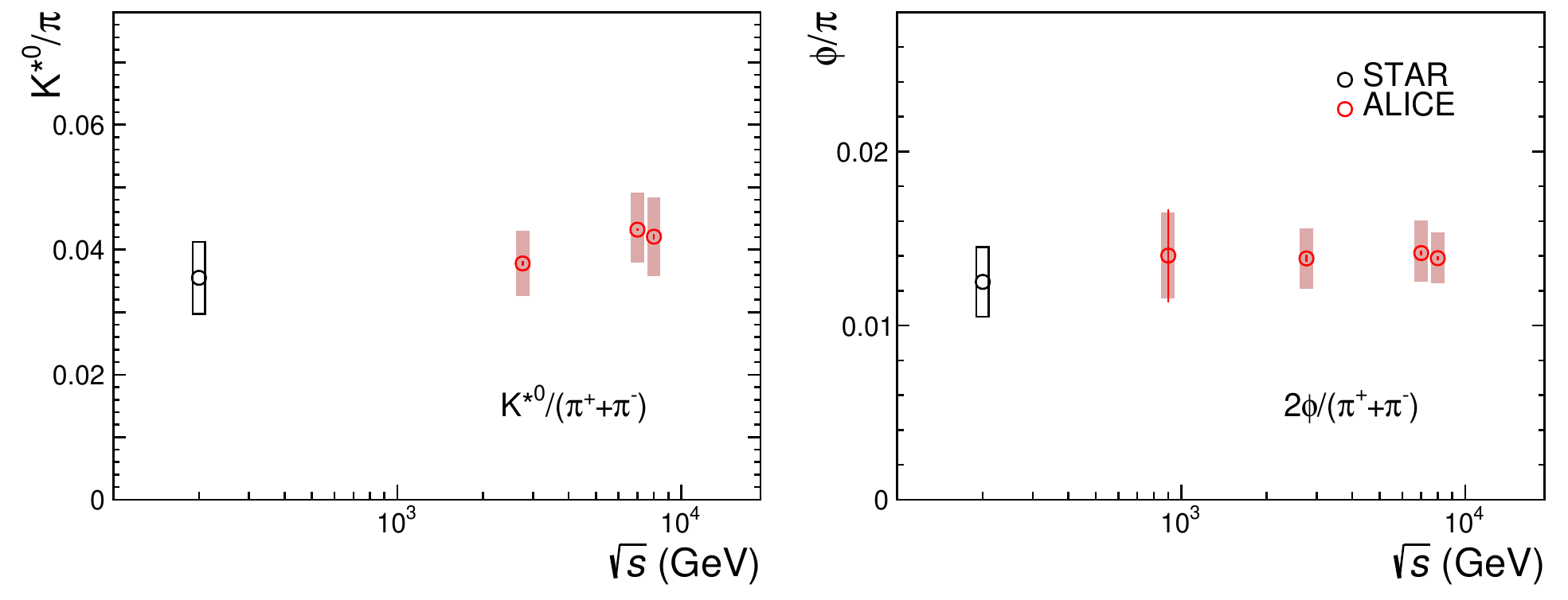}
} 

 \caption{(Color online) Particle ratios of $\ks/\pi$ (left) and $\ph/\pi$ (right) are presented for pp collisions as a function of the collision energy. Bars (when present) represent statistical uncertainties.  Boxes represent the total systematic uncertainties or the total uncertainties for cases when separate statistical uncertainties were not reported.~\cite{STAR_Kstar_200GeV_2005,STAR_phi_200GeV_2005,ALICE_strange_900GeV,ALICE_piKp_900GeV,STAR_phi_2009,Abelev:2014uua,Abelev:2012hy,Adams:2003xp,Adam:2015qaa,Abelev:2014laa}}
    \label{ratio_pp_8:Fig4} 
  \end{center}
\end{figure}

Figure~\ref{ratio_pp_8:Fig4} shows the ratio of the d$N$/d$y$ of $\ks$ ($\ph$) to that of $\pi$ in the left (right) panel, as a function of the collision energy. $\pi$ has no strangeness content, $\ks$ has one unit of strangeness, and $\ph$ is strangeness neutral but contains two strange valence (anti)quarks. It is observed that the $\ks/\pi$ and $\ph/\pi$ ratios are independent of the collision energy within systematic 
uncertainties, which indicates that the chemistry of the system is independent of the energy from the RHIC to LHC energies. This also suggests  
that the strangeness production mechanisms do not depend on energy in inelastic pp collisions at LHC energies.
Figure~\ref{ratio_pp_8:Fig4} and Ref.~\cite{Abelev:2012hy} show that this flat behaviour is observed from RHIC to LHC 
energies and the new result at $\sqrt{s}$ = 8 TeV is in agreement with previous findings. It is worth stressing that this flat behaviour 
is not trivial: since particle yields do in fact increase with the collision energy, the flat ratios are indicative of the fact that the percentage 
increases of d$N$/d$y$ for $\pi$, $\ks$ and $\ph$ as a function of the collision energy are similar from RHIC to LHC.

\begin{figure}[ht]
  \begin{center}
  \resizebox{0.47\textwidth}{!}{
 \includegraphics{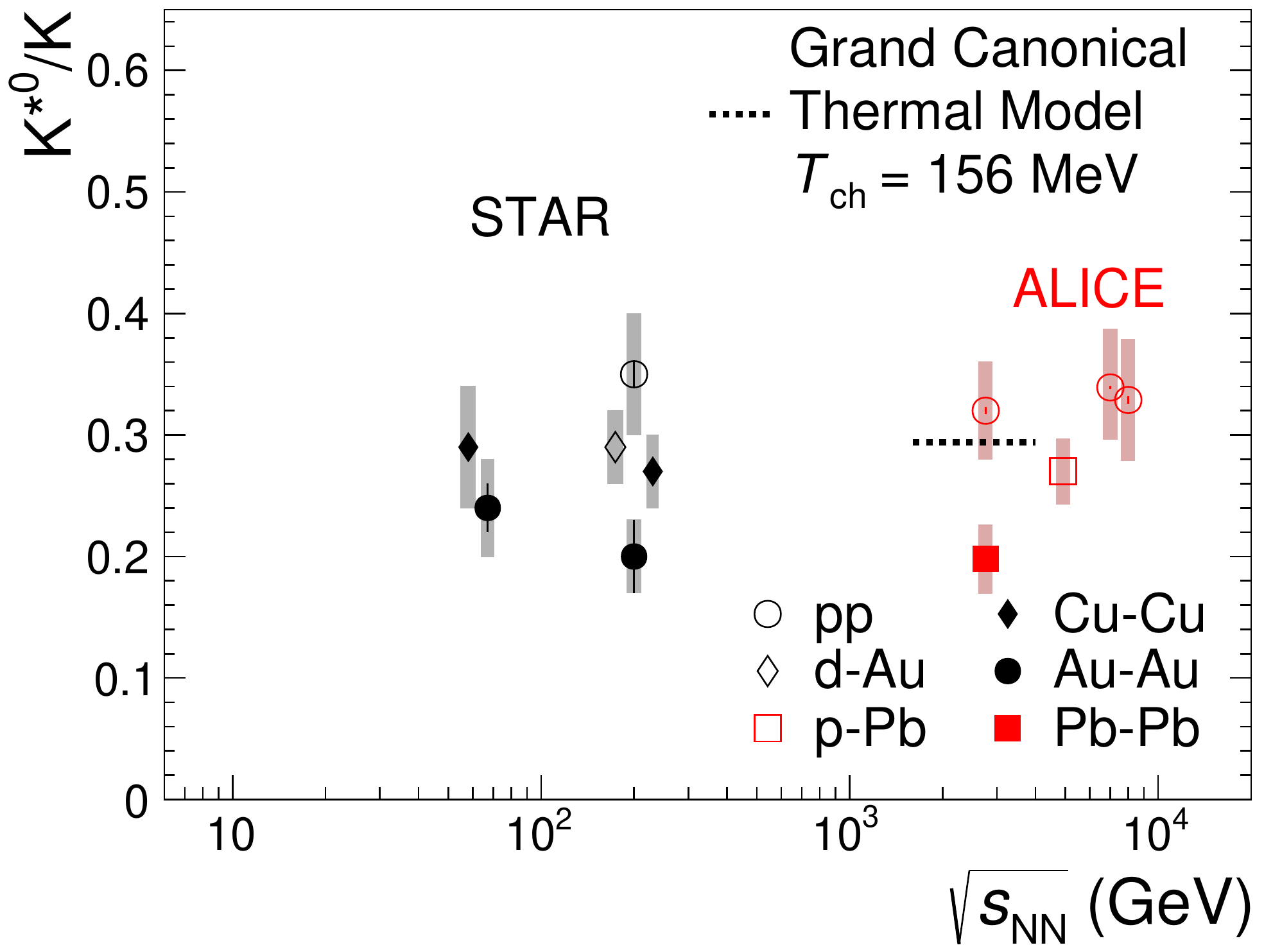}
} 
\resizebox{0.47\textwidth}{!}{
 \includegraphics{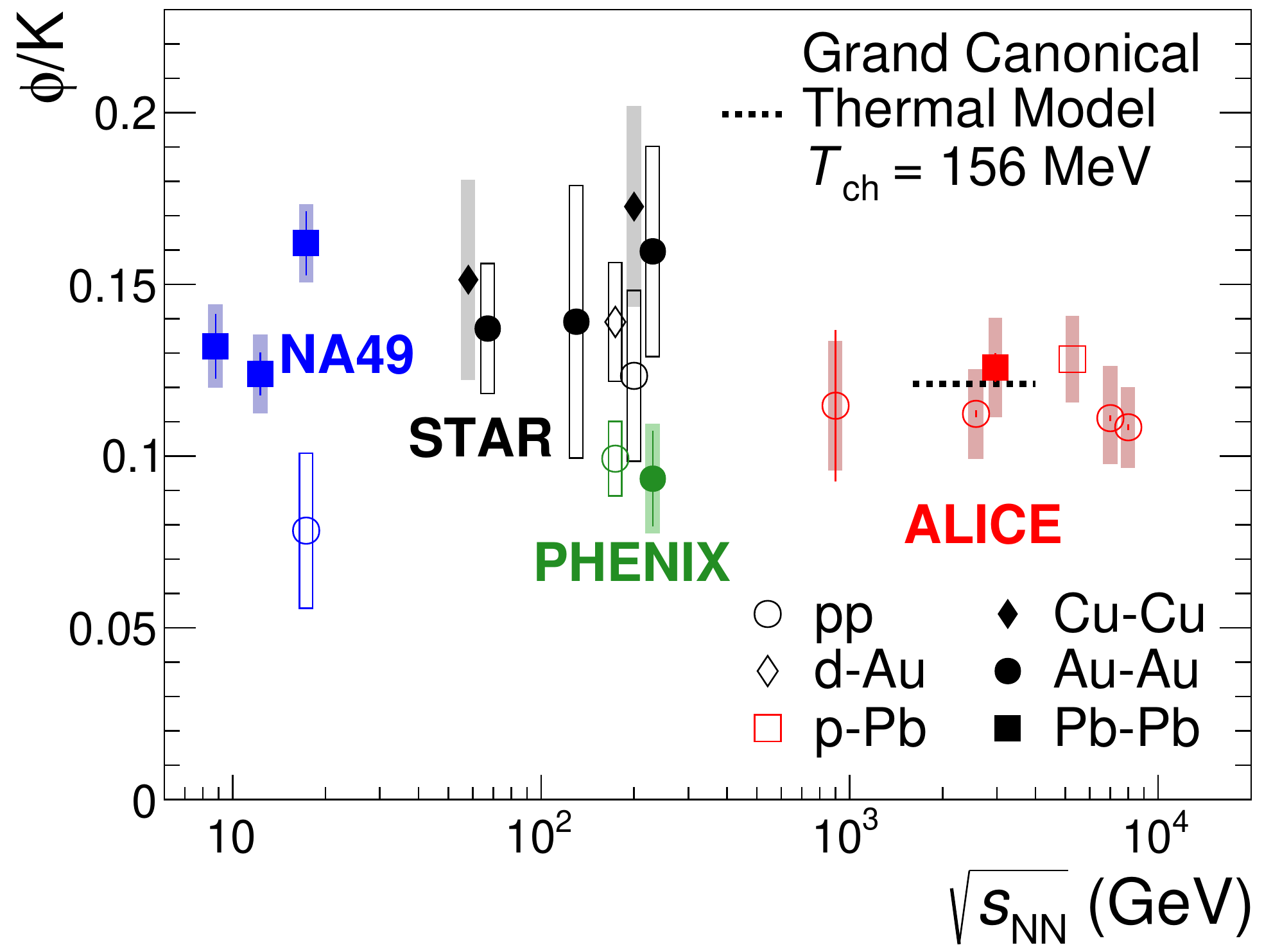}
} 
 \caption{(Color online) Particle ratios of $\ks$/K (left) and $\ph$/K (right) are presented for pp, high-multiplicity p--Pb, central d--Au, and central A--A collisions~\cite{Abelev:2012hy,Abelev:2014laa,STAR_Kstar_200GeV_2005,ALICE_strange_900GeV,PHENIX_phi_AuAu_2005,STAR_Kstar_2011,STAR_phi_2009,NA49_phi_2008,Adam:2015qaa,STAR_resonances_dAu_2008,STAR_phi_200GeV_2005,NA49_phi_2000,NA49_piK_2002,STAR_phi_130GeV,PHENIX_mesons_pp_2011,PHENIX_piKp_pp_2011,ALICE_piKp_900GeV} as a function of the collision energy. Bars (when present) represent statistical uncertainties.  Boxes represent the total systematic uncertainties or the total uncertainties for cases when separate statistical uncertainties were not reported. The value given by a grand-canonical thermal model with a chemical freeze-out temperature of 156 MeV~\cite{Stachel:2013zma} is also shown.} 
    \label{ratio_pp_8:Fig5} 
  \end{center}
\end{figure}

%k*0/k ratio In 
It is interesting to compare the particle ratios, $\ks/\rm{K}$ and $\ph/\rm{K}$ measured in inelastic pp collisions with different collision systems and collision energies in order to understand the production dynamics. In Fig.~\ref{ratio_pp_8:Fig5} the $\ks/\rm{K}$ and $\ph/\rm{K}$ ratios are plotted as a function of center-of-mass energy per nucleon pair for different collision systems. The $\ks/\rm{K}$ and $\ph/\rm{K}$ ratios are independent of the collision energy and of the colliding system. The only exception is the $\ks$ in central nucleus--nucleus collisions; we attribute the suppression of the $\ks/\rm{K}$ ratio to final state effects in the late hadronic stage~\cite{Abelev:2014uua}. The behaviours of these ratios in pp collisions agree with the predictions~\cite{Stachel:2013zma,Abelev:2014uua} of a thermal model in the grand-canonical limit.

\begin{figure}[ht]
 \begin{center}
   \includegraphics[scale=0.4]{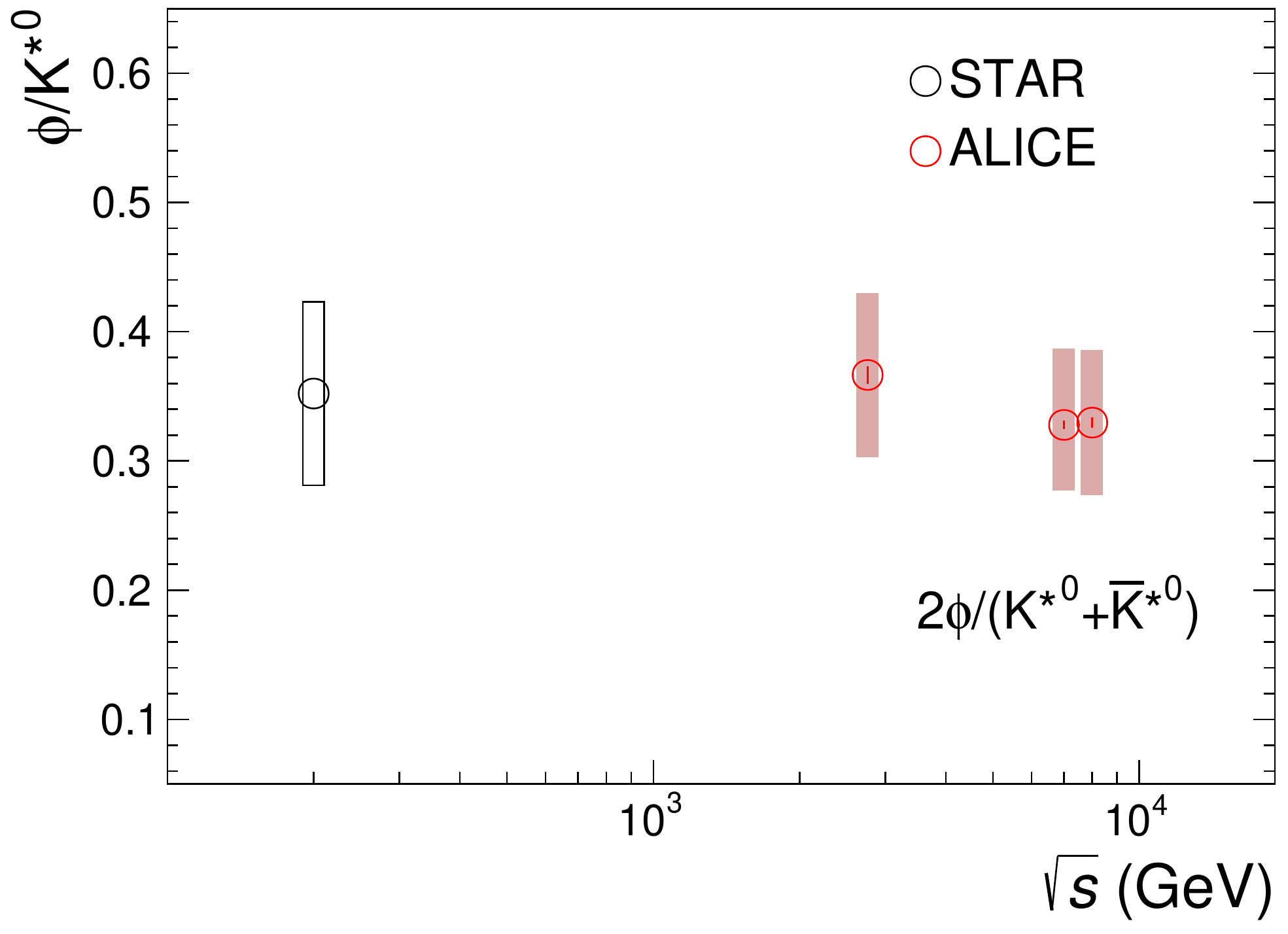}
      \caption{(Color online) Particle ratio $\ph$/  $\ks$ presented for pp collisions~\cite{STAR_Kstar_200GeV_2005,STAR_phi_200GeV_2005,Abelev:2014uua,Abelev:2012hy} as a function 
      of the collision energy. Bars (when present) represent statistical uncertainties.  Boxes represent 
      the total systematic uncertainties or the total uncertainties for cases when separate statistical 
      uncertainties were not reported.}
    \label{pp_8:Fig6}
  \end{center}
\end{figure}

The $\ph$/$\ks$ ratio as a function of center-of-mass energy is plotted  in Fig.~\ref{pp_8:Fig6}.  The ratio seems to be independent of collision energy and appears to follow a behavior expected from thermal production, within experimental uncertainties. 

\subsection{Comparison to models }
QCD-inspired MC event generators like PYTHIA 8~\cite{Skands:2014pea}, PHOJET~\cite{Engel:1995yda,Engel:1994vs} and EPOS-LHC~\cite{Pierog:2013ria} are  used to study multi-particle production, which is predominantly a soft, non-perturbative process. The measurements are compared with the MC model predictions. PYTHIA 8 and PHOJET  use the Lund string fragmentation model~\cite{Andersson:1983ia} for the hadronisation of light and heavy quarks. We compare our data with the Monash 2013 tune~\cite{Skands:2014pea} for PYTHIA 8, which is an updated parameter set for the Lund hadronisation compared to previous tunes.  To describe the non-perturbative phenomena (soft/semi-hard processes), PYTHIA 8 includes multiple parton$-$parton interactions while PHOJET uses the Dual Parton Model~\cite{Capella:1992yb}. For hard scatterings, particle production in both models is based on perturbative QCD and only considers two particle scatterings. For multiple scatterings, the EPOS-LHC model invokes Gribov's Reggeon Field Theory~\cite{Schuler:1993wr}, which features a collective hadronisation via the core-corona mechanism~\cite{Werner:2007bf}. The final state partonic system consists of longitudinal flux tubes which fragment into string segments. The high energy density string segments form the so-called ``core" region, which evolves hydrodynamically to form the bulk part of the system in the final state. The low-density region is known as the ``corona", which expands and breaks via the production of quark-antiquark pairs and hadronises using vacuum string fragmentation. Recent data from the LHC have been used already to tune the EPOS-LHC model~\cite{Pierog:2013ria}.  

\begin{figure}[ht!]
 \begin{center}
   \includegraphics[scale=0.35]{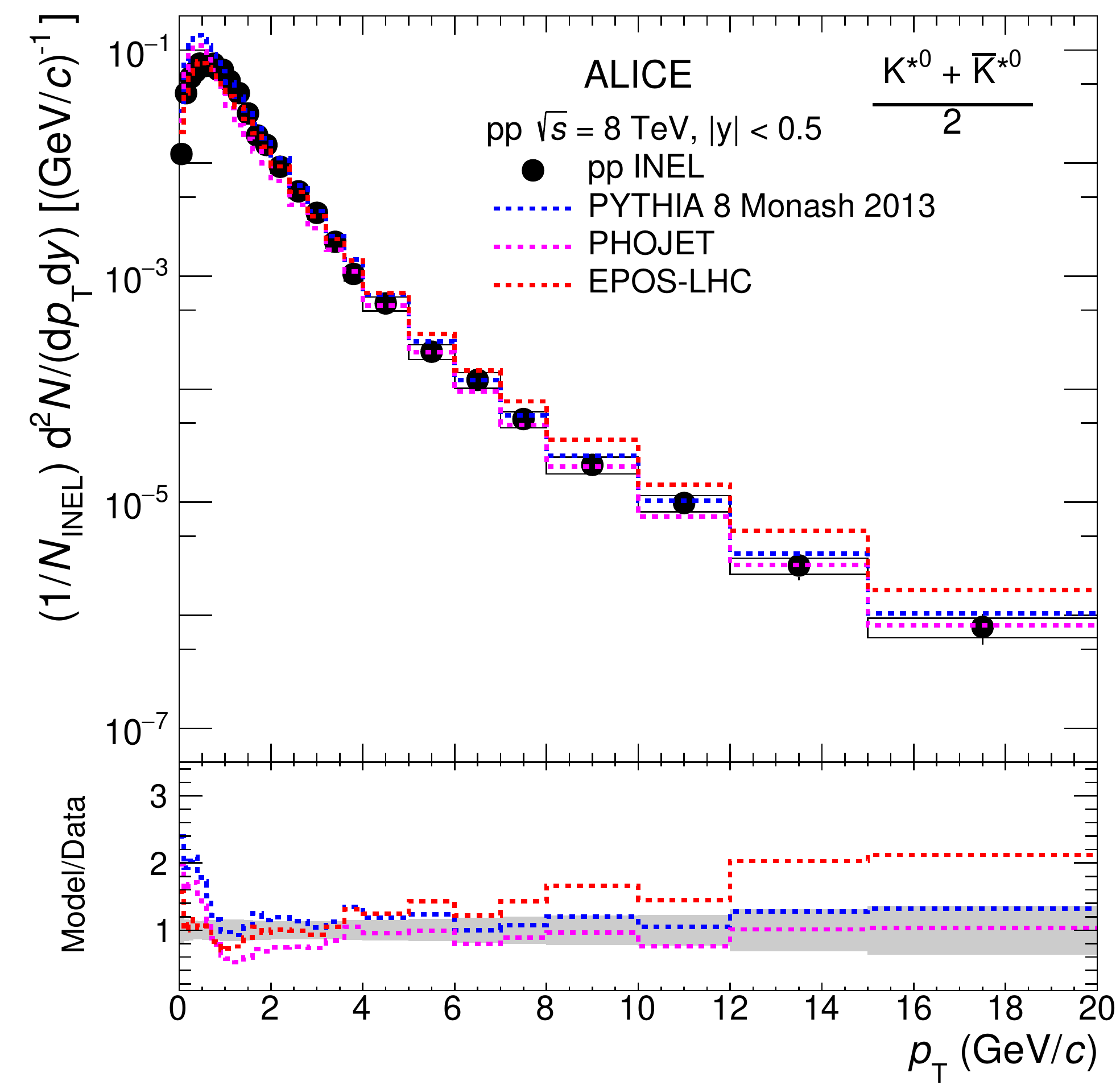}
   \includegraphics[scale=0.35]{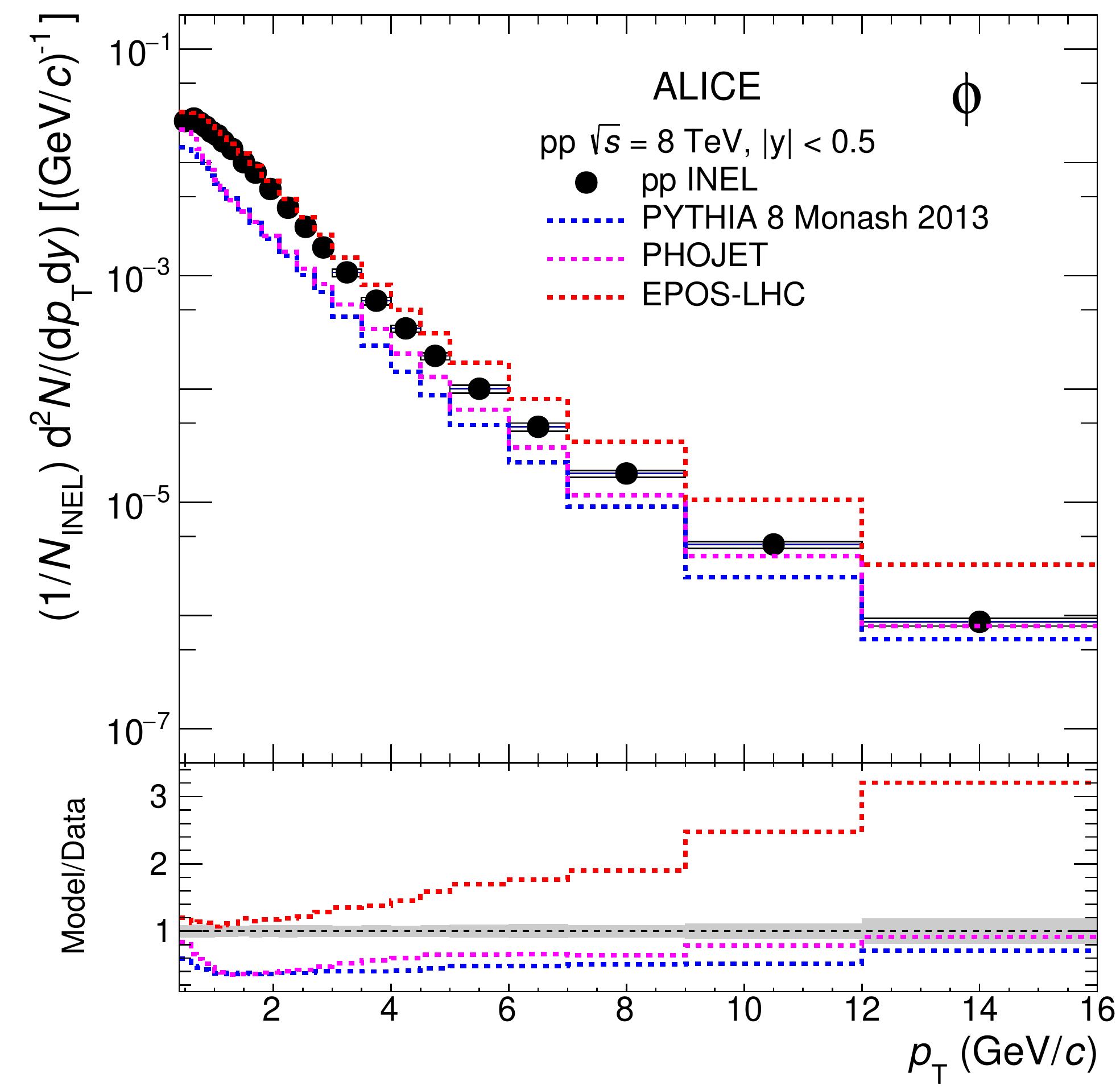}
      \caption{(Color online) Comparison of the $\ks$ (left) and $\ph$ (right) $\pT$ spectra measured in inelastic pp collisions with those obtained from PYTHIA8 (Monash tune)~\cite{Skands:2014pea}, PHOJET~\cite{Engel:1995yda,Engel:1994vs} and EPOS-LHC~\cite{Pierog:2013ria}. The bottom plots show the ratios of the $\pT$ spectra from the models to the measured $\pT$ spectra by ALICE. The total fractional uncertainties of the data are shown as shaded boxes.} 
    \label{pp_8:Fig7}
  \end{center}
\end{figure}

Figure~\ref{pp_8:Fig7} shows a comparison of the $\ks$ (left) and $\ph$ (right) $\pT$  spectra in inelastic pp collisions with PYTHIA8, PHOJET and EPOS-LHC. The bottom panels show the ratios of the $\pT$ spectra from models to the $\pT$ spectra measured by ALICE. The total fractional uncertainties from the real data, including both statistical and systematic uncertainties are shown as shaded boxes. PYTHIA 8 overestimates the $\pT$ spectrum for $\ks$ at very low $\pT$ but describes it in the intermediate-\pT~region and approaches the experimental data at high $\pT$. For the $\ph$ meson, PYTHIA 8 under predicts the yields from the experimental data by about a factor of two. PHOJET has a softer $\pT$ spectrum for $\ks$ and it explains the data above $\pT >$  4 GeV/$c$. For the $\ph$ meson, PHOJET predicts the yields similarly to PYTHIA 8 at low $\pT$, while it approaches the experimental data at higher $\pT$. For the $\ks$, EPOS-LHC describes the $\pT$ spectra at low $\pT$ and overestimates the data above 4 GeV/$c$. For the $\ph$ meson, whereas PYTHIA and PHOJET fail to describe the $\pT$-spectra, the EPOS-LHC model approaches the data at low $\pT$ and deviates monotonically from them with increasing $\pT$.   

\section{Conclusions}
\label{sec5}
Measurements of $\ks$ and $\ph$ production are presented at midrapidity in inelastic pp collisions at $\sqrt{s}$~=~8 TeV in the range 0~$< $ \pT~$<$~20~\gmom~for~$\ks$~and 0.4~$<$ \pT~$< $~16~\gmom~for~$\phi$. Also, updated measurements at $\sqrt{s}$~=~7 TeV are presented, which improve the results previously published in~\cite{Abelev:2012hy}. In comparison to other LHC energies, a hardening of the $\pT$ spectra is observed with increasing collision energy. The $\ks/\pi$ and $\ph/\pi$ ratios are independent of collision energy within systematic uncertainties. This indicates that there is no strangeness enhancement in inelastic pp collisions as the collision energy is increased. Similar behavior is observed for the $\ks/\rm{K}$ and $\ph/\rm{K}$ ratios as a function of collision energy. Also, no energy dependence of the $\ph$/$\ks$ ratio in minimum bias pp collisions at LHC energies is observed, which suggests there is no energy dependence of the chemistry of the system. None of the MC models seem to explain the $\ks$ spectra over the full $\pT$ range whereas PHOJET and PYTHIA describe the data for the intermediate and high-$\pT$ regions. However, the MC models fail to explain the $\pT$ spectra of the $\ph$ meson completely. These pp results will serve as baseline for the measurements in p--Pb and Pb--Pb collisions.

 \newenvironment{acknowledgement}{\relax}{\relax}
 %\end{thebibliography}
\begin{acknowledgement}
\section*{Acknowledgements}
% Version: 2019-06-04

The ALICE Collaboration would like to thank all its engineers and technicians for their invaluable contributions to the construction of the experiment and the CERN accelerator teams for the outstanding performance of the LHC complex.
The ALICE Collaboration gratefully acknowledges the resources and support provided by all Grid centres and the Worldwide LHC Computing Grid (WLCG) collaboration.
The ALICE Collaboration acknowledges the following funding agencies for their support in building and running the ALICE detector:
A. I. Alikhanyan National Science Laboratory (Yerevan Physics Institute) Foundation (ANSL), State Committee of Science and World Federation of Scientists (WFS), Armenia;
Austrian Academy of Sciences, Austrian Science Fund (FWF): [M 2467-N36] and Nationalstiftung f\"{u}r Forschung, Technologie und Entwicklung, Austria;
Ministry of Communications and High Technologies, National Nuclear Research Center, Azerbaijan;
Conselho Nacional de Desenvolvimento Cient\'{\i}fico e Tecnol\'{o}gico (CNPq), Universidade Federal do Rio Grande do Sul (UFRGS), Financiadora de Estudos e Projetos (Finep) and Funda\c{c}\~{a}o de Amparo \`{a} Pesquisa do Estado de S\~{a}o Paulo (FAPESP), Brazil;
Ministry of Science \& Technology of China (MSTC), National Natural Science Foundation of China (NSFC) and Ministry of Education of China (MOEC) , China;
Croatian Science Foundation and Ministry of Science and Education, Croatia;
Centro de Aplicaciones Tecnol\'{o}gicas y Desarrollo Nuclear (CEADEN), Cubaenerg\'{\i}a, Cuba;
Ministry of Education, Youth and Sports of the Czech Republic, Czech Republic;
The Danish Council for Independent Research | Natural Sciences, the Carlsberg Foundation and Danish National Research Foundation (DNRF), Denmark;
Helsinki Institute of Physics (HIP), Finland;
Commissariat \`{a} l'Energie Atomique (CEA), Institut National de Physique Nucl\'{e}aire et de Physique des Particules (IN2P3) and Centre National de la Recherche Scientifique (CNRS) and R\'{e}gion des  Pays de la Loire, France;
Bundesministerium f\"{u}r Bildung und Forschung (BMBF) and GSI Helmholtzzentrum f\"{u}r Schwerionenforschung GmbH, Germany;
General Secretariat for Research and Technology, Ministry of Education, Research and Religions, Greece;
National Research, Development and Innovation Office, Hungary;
Department of Atomic Energy Government of India (DAE), Department of Science and Technology, Government of India (DST), University Grants Commission, Government of India (UGC) and Council of Scientific and Industrial Research (CSIR), India;
Indonesian Institute of Science, Indonesia;
Centro Fermi - Museo Storico della Fisica e Centro Studi e Ricerche Enrico Fermi and Istituto Nazionale di Fisica Nucleare (INFN), Italy;
Institute for Innovative Science and Technology , Nagasaki Institute of Applied Science (IIST), Japan Society for the Promotion of Science (JSPS) KAKENHI and Japanese Ministry of Education, Culture, Sports, Science and Technology (MEXT), Japan;
Consejo Nacional de Ciencia (CONACYT) y Tecnolog\'{i}a, through Fondo de Cooperaci\'{o}n Internacional en Ciencia y Tecnolog\'{i}a (FONCICYT) and Direcci\'{o}n General de Asuntos del Personal Academico (DGAPA), Mexico;
Nederlandse Organisatie voor Wetenschappelijk Onderzoek (NWO), Netherlands;
The Research Council of Norway, Norway;
Commission on Science and Technology for Sustainable Development in the South (COMSATS), Pakistan;
Pontificia Universidad Cat\'{o}lica del Per\'{u}, Peru;
Ministry of Science and Higher Education and National Science Centre, Poland;
Korea Institute of Science and Technology Information and National Research Foundation of Korea (NRF), Republic of Korea;
Ministry of Education and Scientific Research, Institute of Atomic Physics and Ministry of Research and Innovation and Institute of Atomic Physics, Romania;
Joint Institute for Nuclear Research (JINR), Ministry of Education and Science of the Russian Federation, National Research Centre Kurchatov Institute, Russian Science Foundation and Russian Foundation for Basic Research, Russia;
Ministry of Education, Science, Research and Sport of the Slovak Republic, Slovakia;
National Research Foundation of South Africa, South Africa;
Swedish Research Council (VR) and Knut \& Alice Wallenberg Foundation (KAW), Sweden;
European Organization for Nuclear Research, Switzerland;
National Science and Technology Development Agency (NSDTA), Suranaree University of Technology (SUT) and Office of the Higher Education Commission under NRU project of Thailand, Thailand;
Turkish Atomic Energy Agency (TAEK), Turkey;
National Academy of  Sciences of Ukraine, Ukraine;
Science and Technology Facilities Council (STFC), United Kingdom;
National Science Foundation of the United States of America (NSF) and United States Department of Energy, Office of Nuclear Physics (DOE NP), United States of America.
\end{acknowledgement} 
 
\bibliography{refs}{}
\bibliographystyle{utphys}  
    
    \newpage
    \appendix
    \section{The ALICE Collaboration}
    \label{app:collab}
    % Collaboration: CERN-LHC-ALICE
% Generation Date is 2019-Jun-04

% How to use:
%%%%%%%%% appendix with author list
%\appendix
%\section{The ALICE Collaboration}
%\label{app:collab}
%\input{Alice_Authorslist_XXXX-Axx-XX.tex}
\begingroup
\small
\begin{flushleft}
S.~Acharya\Irefn{org141}\And 
D.~Adamov\'{a}\Irefn{org93}\And 
S.P.~Adhya\Irefn{org141}\And 
A.~Adler\Irefn{org73}\And 
J.~Adolfsson\Irefn{org79}\And 
M.M.~Aggarwal\Irefn{org98}\And 
G.~Aglieri Rinella\Irefn{org34}\And 
M.~Agnello\Irefn{org31}\And 
N.~Agrawal\Irefn{org10}\textsuperscript{,}\Irefn{org48}\textsuperscript{,}\Irefn{org53}\And 
Z.~Ahammed\Irefn{org141}
% ...
\And
\renewcommand\labelenumi{\textsuperscript{\theenumi}~}

\section*{Affiliation notes}
\renewcommand\theenumi{\roman{enumi}}
\begin{Authlist}
\item \Adef{org*}Deceased
\item \Adef{orgI}Dipartimento DET del Politecnico di Torino, Turin, Italy
\item \Adef{orgII}M.V. Lomonosov Moscow State University, D.V. Skobeltsyn Institute of Nuclear, Physics, Moscow, Russia
\item \Adef{orgIII}Department of Applied Physics, Aligarh Muslim University, Aligarh, India
\item \Adef{orgIV}Institute of Theoretical Physics, University of Wroclaw, Poland
\end{Authlist}

\section*{Collaboration Institutes}
\renewcommand\theenumi{\arabic{enumi}~}
\begin{Authlist}
\item \Idef{org1}A.I. Alikhanyan National Science Laboratory (Yerevan Physics Institute) Foundation, Yerevan, Armenia
\item \Idef{org2}Bogolyubov Institute for Theoretical Physics, National Academy of Sciences of Ukraine, Kiev, Ukraine
\item \Idef{org3}Bose Institute, Department of Physics  and Centre for Astroparticle Physics and Space Science (CAPSS), Kolkata, India
\item \Idef{org4}Budker Institute for Nuclear Physics, Novosibirsk, Russia
\item \Idef{org5}California Polytechnic State University, San Luis Obispo, California, United States
\item \Idef{org6}Central China Normal University, Wuhan, China
\item \Idef{org7}Centre de Calcul de l'IN2P3, Villeurbanne, Lyon, France
\item \Idef{org8}Centro de Aplicaciones Tecnol\'{o}gicas y Desarrollo Nuclear (CEADEN), Havana, Cuba
\item \Idef{org9}Centro de Investigaci\'{o}n y de Estudios Avanzados (CINVESTAV), Mexico City and M\'{e}rida, Mexico
\item \Idef{org10}Centro Fermi - Museo Storico della Fisica e Centro Studi e Ricerche ``Enrico Fermi', Rome, Italy
\item \Idef{org11}Chicago State University, Chicago, Illinois, United States
\item \Idef{org12}China Institute of Atomic Energy, Beijing, China
\item \Idef{org13}Chonbuk National University, Jeonju, Republic of Korea
\item \Idef{org14}Comenius University Bratislava, Faculty of Mathematics, Physics and Informatics, Bratislava, Slovakia
\item \Idef{org15}COMSATS University Islamabad, Islamabad, Pakistan
\item \Idef{org16}Creighton University, Omaha, Nebraska, United States
\item \Idef{org17}Department of Physics, Aligarh Muslim University, Aligarh, India
\item \Idef{org18}Department of Physics, Pusan National University, Pusan, Republic of Korea
\item \Idef{org19}Department of Physics, Sejong University, Seoul, Republic of Korea
\item \Idef{org20}Department of Physics, University of California, Berkeley, California, United States
\item \Idef{org21}Department of Physics, University of Oslo, Oslo, Norway
\item \Idef{org22}Department of Physics and Technology, University of Bergen, Bergen, Norway
\item \Idef{org23}Dipartimento di Fisica dell'Universit\`{a} 'La Sapienza' and Sezione INFN, Rome, Italy
\item \Idef{org24}Dipartimento di Fisica dell'Universit\`{a} and Sezione INFN, Cagliari, Italy
\item \Idef{org25}Dipartimento di Fisica dell'Universit\`{a} and Sezione INFN, Trieste, Italy
\item \Idef{org26}Dipartimento di Fisica dell'Universit\`{a} and Sezione INFN, Turin, Italy
\item \Idef{org27}Dipartimento di Fisica e Astronomia dell'Universit\`{a} and Sezione INFN, Bologna, Italy
\item \Idef{org28}Dipartimento di Fisica e Astronomia dell'Universit\`{a} and Sezione INFN, Catania, Italy
\item \Idef{org29}Dipartimento di Fisica e Astronomia dell'Universit\`{a} and Sezione INFN, Padova, Italy
\item \Idef{org30}Dipartimento di Fisica `E.R.~Caianiello' dell'Universit\`{a} and Gruppo Collegato INFN, Salerno, Italy
\item \Idef{org31}Dipartimento DISAT del Politecnico and Sezione INFN, Turin, Italy
\item \Idef{org32}Dipartimento di Scienze e Innovazione Tecnologica dell'Universit\`{a} del Piemonte Orientale and INFN Sezione di Torino, Alessandria, Italy
\item \Idef{org33}Dipartimento Interateneo di Fisica `M.~Merlin' and Sezione INFN, Bari, Italy
\item \Idef{org34}European Organization for Nuclear Research (CERN), Geneva, Switzerland
\item \Idef{org35}Faculty of Electrical Engineering, Mechanical Engineering and Naval Architecture, University of Split, Split, Croatia
\item \Idef{org36}Faculty of Engineering and Science, Western Norway University of Applied Sciences, Bergen, Norway
\item \Idef{org37}Faculty of Nuclear Sciences and Physical Engineering, Czech Technical University in Prague, Prague, Czech Republic
\item \Idef{org38}Faculty of Science, P.J.~\v{S}af\'{a}rik University, Ko\v{s}ice, Slovakia
\item \Idef{org39}Frankfurt Institute for Advanced Studies, Johann Wolfgang Goethe-Universit\"{a}t Frankfurt, Frankfurt, Germany
\item \Idef{org40}Gangneung-Wonju National University, Gangneung, Republic of Korea
\item \Idef{org41}Gauhati University, Department of Physics, Guwahati, India
\item \Idef{org42}Helmholtz-Institut f\"{u}r Strahlen- und Kernphysik, Rheinische Friedrich-Wilhelms-Universit\"{a}t Bonn, Bonn, Germany
\item \Idef{org43}Helsinki Institute of Physics (HIP), Helsinki, Finland
\item \Idef{org44}High Energy Physics Group,  Universidad Aut\'{o}noma de Puebla, Puebla, Mexico
\item \Idef{org45}Hiroshima University, Hiroshima, Japan
\item \Idef{org46}Hochschule Worms, Zentrum  f\"{u}r Technologietransfer und Telekommunikation (ZTT), Worms, Germany
\item \Idef{org47}Horia Hulubei National Institute of Physics and Nuclear Engineering, Bucharest, Romania
\item \Idef{org48}Indian Institute of Technology Bombay (IIT), Mumbai, India
\item \Idef{org49}Indian Institute of Technology Indore, Indore, India
\item \Idef{org50}Indonesian Institute of Sciences, Jakarta, Indonesia
\item \Idef{org51}INFN, Laboratori Nazionali di Frascati, Frascati, Italy
\item \Idef{org52}INFN, Sezione di Bari, Bari, Italy
\item \Idef{org53}INFN, Sezione di Bologna, Bologna, Italy
\item \Idef{org54}INFN, Sezione di Cagliari, Cagliari, Italy
\item \Idef{org55}INFN, Sezione di Catania, Catania, Italy
\item \Idef{org56}INFN, Sezione di Padova, Padova, Italy
\item \Idef{org57}INFN, Sezione di Roma, Rome, Italy
\item \Idef{org58}INFN, Sezione di Torino, Turin, Italy
\item \Idef{org59}INFN, Sezione di Trieste, Trieste, Italy
\item \Idef{org60}Inha University, Incheon, Republic of Korea
\item \Idef{org61}Institut de Physique Nucl\'{e}aire d'Orsay (IPNO), Institut National de Physique Nucl\'{e}aire et de Physique des Particules (IN2P3/CNRS), Universit\'{e} de Paris-Sud, Universit\'{e} Paris-Saclay, Orsay, France
\item \Idef{org62}Institute for Nuclear Research, Academy of Sciences, Moscow, Russia
\item \Idef{org63}Institute for Subatomic Physics, Utrecht University/Nikhef, Utrecht, Netherlands
\item \Idef{org64}Institute of Experimental Physics, Slovak Academy of Sciences, Ko\v{s}ice, Slovakia
\item \Idef{org65}Institute of Physics, Homi Bhabha National Institute, Bhubaneswar, India
\item \Idef{org66}Institute of Physics of the Czech Academy of Sciences, Prague, Czech Republic
\item \Idef{org67}Institute of Space Science (ISS), Bucharest, Romania
\item \Idef{org68}Institut f\"{u}r Kernphysik, Johann Wolfgang Goethe-Universit\"{a}t Frankfurt, Frankfurt, Germany
\item \Idef{org69}Instituto de Ciencias Nucleares, Universidad Nacional Aut\'{o}noma de M\'{e}xico, Mexico City, Mexico
\item \Idef{org70}Instituto de F\'{i}sica, Universidade Federal do Rio Grande do Sul (UFRGS), Porto Alegre, Brazil
\item \Idef{org71}Instituto de F\'{\i}sica, Universidad Nacional Aut\'{o}noma de M\'{e}xico, Mexico City, Mexico
\item \Idef{org72}iThemba LABS, National Research Foundation, Somerset West, South Africa
\item \Idef{org73}Johann-Wolfgang-Goethe Universit\"{a}t Frankfurt Institut f\"{u}r Informatik, Fachbereich Informatik und Mathematik, Frankfurt, Germany
\item \Idef{org74}Joint Institute for Nuclear Research (JINR), Dubna, Russia
\item \Idef{org75}Korea Institute of Science and Technology Information, Daejeon, Republic of Korea
\item \Idef{org76}KTO Karatay University, Konya, Turkey
\item \Idef{org77}Laboratoire de Physique Subatomique et de Cosmologie, Universit\'{e} Grenoble-Alpes, CNRS-IN2P3, Grenoble, France
\item \Idef{org78}Lawrence Berkeley National Laboratory, Berkeley, California, United States
\item \Idef{org79}Lund University Department of Physics, Division of Particle Physics, Lund, Sweden
\item \Idef{org80}Nagasaki Institute of Applied Science, Nagasaki, Japan
\item \Idef{org81}Nara Women{'}s University (NWU), Nara, Japan
\item \Idef{org82}National and Kapodistrian University of Athens, School of Science, Department of Physics , Athens, Greece
\item \Idef{org83}National Centre for Nuclear Research, Warsaw, Poland
\item \Idef{org84}National Institute of Science Education and Research, Homi Bhabha National Institute, Jatni, India
\item \Idef{org85}National Nuclear Research Center, Baku, Azerbaijan
\item \Idef{org86}National Research Centre Kurchatov Institute, Moscow, Russia
\item \Idef{org87}Niels Bohr Institute, University of Copenhagen, Copenhagen, Denmark
\item \Idef{org88}Nikhef, National institute for subatomic physics, Amsterdam, Netherlands
\item \Idef{org89}NRC Kurchatov Institute IHEP, Protvino, Russia
\item \Idef{org90}NRC Kurchatov Institute  - ITEP, Moscow, Russia
\item \Idef{org91}NRNU Moscow Engineering Physics Institute, Moscow, Russia
\item \Idef{org92}Nuclear Physics Group, STFC Daresbury Laboratory, Daresbury, United Kingdom
\item \Idef{org93}Nuclear Physics Institute of the Czech Academy of Sciences, \v{R}e\v{z} u Prahy, Czech Republic
\item \Idef{org94}Oak Ridge National Laboratory, Oak Ridge, Tennessee, United States
\item \Idef{org95}Ohio State University, Columbus, Ohio, United States
\item \Idef{org96}Petersburg Nuclear Physics Institute, Gatchina, Russia
\item \Idef{org97}Physics department, Faculty of science, University of Zagreb, Zagreb, Croatia
\item \Idef{org98}Physics Department, Panjab University, Chandigarh, India
\item \Idef{org99}Physics Department, University of Jammu, Jammu, India
\item \Idef{org100}Physics Department, University of Rajasthan, Jaipur, India
\item \Idef{org101}Physikalisches Institut, Eberhard-Karls-Universit\"{a}t T\"{u}bingen, T\"{u}bingen, Germany
\item \Idef{org102}Physikalisches Institut, Ruprecht-Karls-Universit\"{a}t Heidelberg, Heidelberg, Germany
\item \Idef{org103}Physik Department, Technische Universit\"{a}t M\"{u}nchen, Munich, Germany
\item \Idef{org104}Politecnico di Bari, Bari, Italy
\item \Idef{org105}Research Division and ExtreMe Matter Institute EMMI, GSI Helmholtzzentrum f\"ur Schwerionenforschung GmbH, Darmstadt, Germany
\item \Idef{org106}Rudjer Bo\v{s}kovi\'{c} Institute, Zagreb, Croatia
\item \Idef{org107}Russian Federal Nuclear Center (VNIIEF), Sarov, Russia
\item \Idef{org108}Saha Institute of Nuclear Physics, Homi Bhabha National Institute, Kolkata, India
\item \Idef{org109}School of Physics and Astronomy, University of Birmingham, Birmingham, United Kingdom
\item \Idef{org110}Secci\'{o}n F\'{\i}sica, Departamento de Ciencias, Pontificia Universidad Cat\'{o}lica del Per\'{u}, Lima, Peru
\item \Idef{org111}Shanghai Institute of Applied Physics, Shanghai, China
\item \Idef{org112}St. Petersburg State University, St. Petersburg, Russia
\item \Idef{org113}Stefan Meyer Institut f\"{u}r Subatomare Physik (SMI), Vienna, Austria
\item \Idef{org114}SUBATECH, IMT Atlantique, Universit\'{e} de Nantes, CNRS-IN2P3, Nantes, France
\item \Idef{org115}Suranaree University of Technology, Nakhon Ratchasima, Thailand
\item \Idef{org116}Technical University of Ko\v{s}ice, Ko\v{s}ice, Slovakia
\item \Idef{org117}Technische Universit\"{a}t M\"{u}nchen, Excellence Cluster 'Universe', Munich, Germany
\item \Idef{org118}The Henryk Niewodniczanski Institute of Nuclear Physics, Polish Academy of Sciences, Cracow, Poland
\item \Idef{org119}The University of Texas at Austin, Austin, Texas, United States
\item \Idef{org120}Universidad Aut\'{o}noma de Sinaloa, Culiac\'{a}n, Mexico
\item \Idef{org121}Universidade de S\~{a}o Paulo (USP), S\~{a}o Paulo, Brazil
\item \Idef{org122}Universidade Estadual de Campinas (UNICAMP), Campinas, Brazil
\item \Idef{org123}Universidade Federal do ABC, Santo Andre, Brazil
\item \Idef{org124}University of Cape Town, Cape Town, South Africa
\item \Idef{org125}University of Houston, Houston, Texas, United States
\item \Idef{org126}University of Jyv\"{a}skyl\"{a}, Jyv\"{a}skyl\"{a}, Finland
\item \Idef{org127}University of Liverpool, Liverpool, United Kingdom
\item \Idef{org128}University of Science and Techonology of China, Hefei, China
\item \Idef{org129}University of South-Eastern Norway, Tonsberg, Norway
\item \Idef{org130}University of Tennessee, Knoxville, Tennessee, United States
\item \Idef{org131}University of the Witwatersrand, Johannesburg, South Africa
\item \Idef{org132}University of Tokyo, Tokyo, Japan
\item \Idef{org133}University of Tsukuba, Tsukuba, Japan
\item \Idef{org134}Universit\'{e} Clermont Auvergne, CNRS/IN2P3, LPC, Clermont-Ferrand, France
\item \Idef{org135}Universit\'{e} de Lyon, Universit\'{e} Lyon 1, CNRS/IN2P3, IPN-Lyon, Villeurbanne, Lyon, France
\item \Idef{org136}Universit\'{e} de Strasbourg, CNRS, IPHC UMR 7178, F-67000 Strasbourg, France, Strasbourg, France
\item \Idef{org137}Universit\'{e} Paris-Saclay Centre d'Etudes de Saclay (CEA), IRFU, D\'{e}partment de Physique Nucl\'{e}aire (DPhN), Saclay, France
\item \Idef{org138}Universit\`{a} degli Studi di Foggia, Foggia, Italy
\item \Idef{org139}Universit\`{a} degli Studi di Pavia, Pavia, Italy
\item \Idef{org140}Universit\`{a} di Brescia, Brescia, Italy
\item \Idef{org141}Variable Energy Cyclotron Centre, Homi Bhabha National Institute, Kolkata, India
\item \Idef{org142}Warsaw University of Technology, Warsaw, Poland
\item \Idef{org143}Wayne State University, Detroit, Michigan, United States
\item \Idef{org144}Westf\"{a}lische Wilhelms-Universit\"{a}t M\"{u}nster, Institut f\"{u}r Kernphysik, M\"{u}nster, Germany
\item \Idef{org145}Wigner Research Centre for Physics, Hungarian Academy of Sciences, Budapest, Hungary
\item \Idef{org146}Yale University, New Haven, Connecticut, United States
\item \Idef{org147}Yonsei University, Seoul, Republic of Korea
\end{Authlist}
\endgroup
  %%%%%%% done by webmaster team                                                                            
\end{document}